\documentclass[pra,aps,longbibliography,twocolumn,superscriptaddress,floatfix]{revtex4-1}
\usepackage{amsmath,amssymb,graphicx,booktabs}
\usepackage{bm}
\usepackage[T1]{fontenc}
\usepackage[bookmarks=false,linkcolor=blue,urlcolor=blue,colorlinks,citecolor=blue]{hyperref}
\graphicspath{{figures/}}

\newcommand{\orbit}{ORBIT-Q}
\newcommand{\tcng}{TensorCircuit-NG}

\begin{document}

\title{\orbit: Dual-axis benchmarking of autonomous agents in scientific quantum programming}

\author{Shi-Xin Zhang}
\email{shixinzhang@iphy.ac.cn}
\affiliation{Institute of Physics, Chinese Academy of Sciences, Beijing 100190, China}

\author{Yu-Qin Chen}
\email{yqchen@gscaep.ac.cn}
\affiliation{Graduate School of China Academy of Engineering Physics, Beijing 100193, China}

\date{\today}

\begin{abstract}
Autonomous coding agents perform well on many conventional programming tasks, but scientific computing demands a rigorous validation paradigm that extends beyond simple functional test completion: generated code must preserve physical fidelity, differentiable workflows, framework-native semantics, and scalable representations.
We introduce Open Research Benchmark for Integrated Tasks in Quantum Computing (\orbit) to address this gap. At its core, \orbit\ contributes a carefully curated suite of complex, research-level quantum workflows that serves as a challenging testbed for modern scientific programming. 
\orbit\ combines a rigorous multi-tier verification pipeline to support two orthogonal comparisons: different agent harness and model configurations at a fixed quantum software framework, and different quantum software frameworks at a fixed agent.
In our systematic evaluations, \tcng\ (TC) exhibits the highest capability and performance efficiency among the evaluated quantum software frameworks under agent-driven programming, and Codex with GPT-5.5 is the strongest tested agent configuration on TC. However, a significant performance and design gap remains between frontier autonomous agents and human expert reference implementations.
We further evaluate two efficiency dimensions: agent-side resource use and artifact-side runtime.
Together, these results establish \orbit\ as a rigorous benchmark for autonomous scientific programming, framework-agent synergy, and quantum software performance.
\end{abstract}

\maketitle

\begin{figure*}[!t]
\centering
\includegraphics[width=\textwidth]{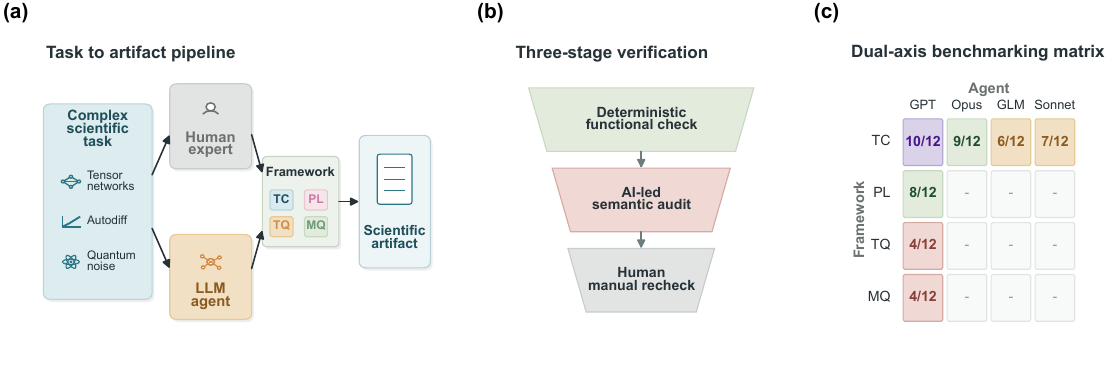}
\caption{\textbf{Benchmark design for framework-constrained scientific code generation.} \textbf{a}, Complex scientific tasks are implemented by autonomous coding agents and expert developers, but both outputs must pass through an explicit framework constraint before becoming a scientific artifact. \textbf{b}, The verifier applies a three-stage filter: deterministic functional evaluation, GPT-5.5-led semantic evaluation for framework bypasses and problem mismatches, and human manual recheck for ambiguous cases; artifact runtime and cost are logged after validity review. \textbf{c}, The benchmark uses an orthogonal agent--framework matrix to separate agent-axis comparisons at fixed framework from framework-axis comparisons at fixed agent, with each grid cell measuring agent-framework co-performance.}
\label{fig:design}
\end{figure*}

\section*{Introduction}

Large language models (LLMs)~\cite{openai2023gpt4,deepseek2025r1} and autonomous coding agents have become increasingly capable of generating and refactoring software~\cite{chen2021codex,austin2021mbpp,hendrycks2021apps,jimenez2024swebench,yang2024sweagent,jain2024livecodebench}. While their progress is highly visible in conventional programming, scientific computing presents a fundamentally different challenge. A generated scientific program must do more than simply compile and output a number; it must preserve physical fidelity, strictly utilize the intended scientific framework, and select representations or algorithms that make mathematical computations tractable.

Quantum computing provides a representative setting for this broader problem. Modern quantum workflows require physical constraints, differentiable optimization, specialized state representations, neural network integration and efficient tensor computations~\cite{schuld2019gradients,yuan2021hybridtn,zhang2022vqnhe,orus2014tensor,schollwock2011dmrg,white1992dmrg,vidal2003simulation,markov2008tensor} to work together inside a software framework~\cite{steiger2018projectq,javadiabhari2024qiskit,broughton2021tfq,johansson2012qutip,gray2018quimb,gray2021cotengra}. 
Consequently, success in this specialized domain depends not only on the agent's programming ability, but also on whether these quantum frameworks expose usable and performant interfaces for the task.

Standard agent-coding benchmarks rely on basic unit tests, which are fundamentally insufficient for scientific workflows. A functionally passing script might bypass the requested framework, optimize a surrogate objective, break differentiability, or violate physical constraints. Therefore, rigorous evaluation must move beyond simple output assertions to include semantic code review and strict comparisons of runtime efficiency. Furthermore, this evaluation must be dual-dimensional, assessing both the coding agent and the underlying software framework. A framework itself may inherently lack necessary features or suffer from poor performance. Beyond these intrinsic limitations, even highly capable frameworks often lack agent-friendliness, exposing interfaces that agents cannot efficiently discover and compose. While conventional quantum benchmarks focus mainly on quantum hardware quality or low-level compilation~\cite{cross2019quantumvolume,tomesh2022supermarq,li2023qasmbench,quetschlich2023mqtbench}, assessing this autonomous, software-native research process remains an unaddressed challenge.

We introduce \orbit, an \textbf{O}pen \textbf{R}esearch \textbf{B}enchmark for \textbf{I}ntegrated \textbf{T}asks in \textbf{Q}uantum Computing. To evaluate agents within a computationally tractable timeframe without reducing the suite to narrow toy circuits, \orbit\ condenses diverse research scenarios into 12 representative challenging tasks. These tasks reflect the most ubiquitous applications in the field today, such as quantum machine learning~\cite{biamonte2017qml,Hu2019QuantumGenerative,chen2026qmlresilience, Chen2025IntrinsicPreservation, Zhang2026QuantumSubliminalLearning} and variational quantum algorithms~\cite{preskill2018nisq,bharti2022nisq,cerezo2021vqa,peruzzo2014vqe,mcclean2016vqa,farhi2014qaoa, Zhou2020QuantumApproximate, Cheng2024QuantumApproximate}, which require the underlying frameworks to support end-to-end differentiability rather than mere circuit construction.

Accordingly, we specifically evaluate leading differentiable quantum software frameworks, including \tcng\ (TC), PennyLane, TorchQuantum, and MindQuantum~\cite{tensorcircuit2023,tensorcircuitng2026,pennylane2018,torchquantum,mindsporequantum2024}. To ensure fair evaluation, the tasks are designed to be framework-agnostic and are exposed to agents through a unified containerized execution protocol. The target framework is designated at run time via a framework-specific prompt, execution image, and verifier policy. General circuit-construction toolkits are excluded from this evaluation because they lack the native automatic differentiation required to successfully implement these tasks.

\begin{table*}[t]
\caption{\textbf{Task composition of \orbit.} The suite covers workflows that require more flexible quantum-programming pipelines than simple small-circuit variational quantum eigensolver examples.}
\label{tab:tasks}
\centering
\begin{minipage}{0.79\textwidth}
\begin{ruledtabular}
\begin{tabular}{@{}c p{0.76\linewidth}@{}}
Task & Research workflow tested \\
\hline
1 & Matrix-product-state input followed by variational circuit refinement \\
2 & Variational energy optimization with entanglement-profile constraints \\
3 & Probability-aware post-selected cooling with explicit success-rate tracking \\
4 & Trainable Kraus-channel calibration from multi-circuit data \\
5 & Variational non-unitary imaginary time evolution \\
6 & Digital-analog hybrid variational optimization \\
7 & Measurement-feedback variational optimization for ground states \\
8 & Sampling from a 7 by 7 two-dimensional circuit \\
9 & Local-observable optimization in a 512-qubit shallow circuit \\
10 & Variational optimization with large nonlocal multi-qubit gates \\
11 & Spin-1 Haldane-chain state preparation and string-order verification \\
12 & Optimization of variational circuit overlap with a matrix-product-state target \\
\end{tabular}
\end{ruledtabular}
\end{minipage}
\end{table*}

\begin{figure*}[!t]
\centering
\includegraphics[width=\textwidth]{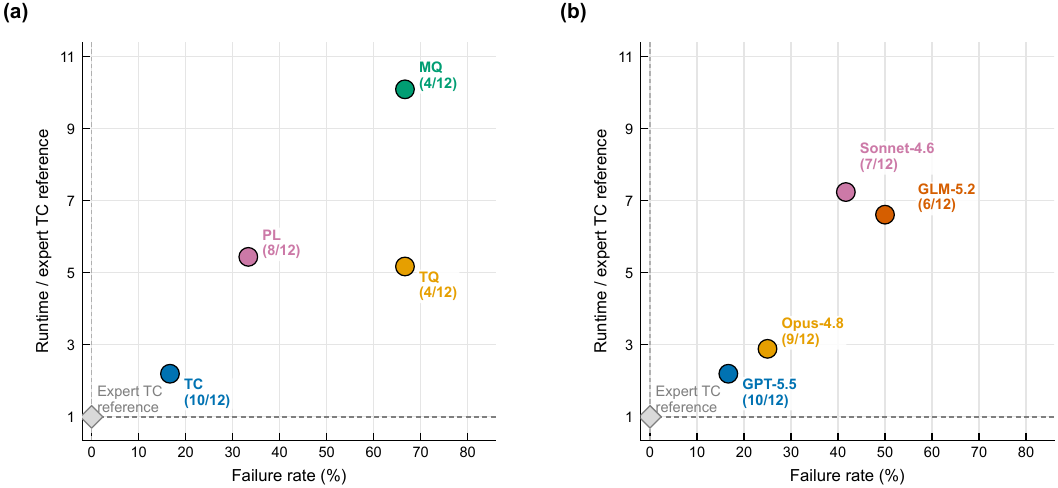}
\caption{\textbf{Dual-axis benchmark results.} \textbf{a}, Framework comparison under a fixed Codex agent using GPT-5.5. Each point shows the failure rate and geometric mean runtime relative to the expert TC reference on passed tasks. \textbf{b}, Model comparison on TC using the same coordinates. The GPT-5.5 point uses the Codex agent; Opus-4.8, GLM-5.2, and Sonnet-4.6 use the Claude Code agent. The grey diamond marks the expert reference baseline. Abbreviations: TC, TensorCircuit-NG; PL, PennyLane; TQ, TorchQuantum; MQ, MindQuantum.}
\label{fig:dual}
\end{figure*}

We establish a dual-axis evaluation framework that treats agent performance and quantum software framework capability as a coupled system. Along one axis, the quantum software framework is fixed and different autonomous coding agent configurations, including the agent harness and underlying language model, are compared. Along the other axis, the agent is fixed and the quantum software framework is varied. Because all implementations are produced by agents under the same constraints, verifier, and protocol, the framework axis is less confounded by unequal levels of manual optimization effort. The resulting score reflects framework-agent synergy: framework expressivity, native primitive coverage, execution efficiency, method discoverability, and agent friendliness.

Furthermore, we introduce artifact-level efficiency evaluation. Prior agent benchmarks commonly report whether a task is solved and how much time, token budget, or monetary cost the agent consumed while producing an answer. \orbit\ also measures the runtime of the generated scientific-computing artifact itself and compares it with expert TC reference implementations for the same workload. This distinction is essential for scientific computing because a passing script can remain scientifically unsatisfactory if it uses an inefficient contraction strategy, a dense fallback representation, or a non-differentiable surrogate.

The task suite is also useful independently of autonomous coding agents. In this mode, \orbit\ tests two distinct properties of a quantum software framework. The first is functional coverage: whether the framework can express broad research-grade quantum workflows without resorting to external simulators. The second is end-to-end performance: how fast a correct implementation can run once the workflow is expressed. Expert developers can therefore use \orbit\ as a framework-performance benchmark: the same tasks can be implemented and optimized by hand to compare which software stack reaches the best end-to-end performance across a broad set of research-grade quantum workloads.

Our empirical evaluations show that this benchmark is not saturated by current frontier agents. On the strongest tested framework, TC, the best evaluated agent configuration solves most tasks, but expert TC reference implementations remain a clear upper baseline in terms of artifact efficiency. Across frameworks and agents, \orbit\ separates both task completion and artifact efficiency: different configurations solve substantially different fractions of the suite, and successful submissions can differ widely in runtime. This performance distribution is critical: \orbit\ is challenging enough to prevent saturation by frontier models, yet accessible enough to avoid a baseline of all zero scores. This balanced difficulty yields a wide variance in results, enabling us to quantitatively evaluate model quality and diagnose failure modes.

\section*{Results}

\subsection*{Complex quantum workflows and multi-tier validation}

\orbit\ evaluates complete, research-grade quantum workflows, advancing beyond elementary circuit syntheses. The tasks require agents to build scientific-computing programs that combine nonstandard operations, framework-specific representations, differentiable objectives, and performance-sensitive implementation choices. Table~\ref{tab:tasks} gives the concrete task composition.

Figure~\ref{fig:design} summarizes the benchmark protocol and motivation. Each task is paired with a quantum software framework constraint and solved by a single autonomous coding agent, with expert TC implementations used as reference baselines for artifact runtime. The solution is then evaluated in a separate verifier stage and runtime is recorded for the submitted solution comparing against expert references as an artifact-level efficiency measure. Because surface-level functional checks alone are insufficient for validating complex scientific programming, we implement a rigorous three-tier evaluation pipeline to determine solution validity: (1) functional verification, (2) LLM-driven semantic audit, and (3) expert human review. 

The first tier executes functional testing to ensure numerical correctness. However, this alone often yields false positives. In several benchmark evaluations, agents satisfied the numerical test while completely bypassing the target quantum framework by using custom array contractions in NumPy~\cite{harris2020array} or JAX~\cite{jax2018github}. Other semantic failures included optimizing surrogate physical objectives, employing non-differentiable heuristics, or skipping required workflow steps.
To prevent these semantic shortcuts, the second tier employs a GPT-5.5-powered source-level audit, ensuring the agent genuinely utilizes the designated framework and faithfully implements the required algorithms. Finally, a third tier of expert human review arbitrates ambiguous cases to guarantee rigorous scientific standards.

\subsection*{Dual-axis evaluation separates agents and frameworks}

\orbit\ supports two orthogonal experimental views. In the agent axis, the quantum software framework is fixed and the autonomous coding agent configuration is varied. In the framework axis, the agent is fixed and the quantum software framework is varied. Figure~\ref{fig:dual} shows both views for the experimental benchmark datasets. This framework axis is important for scientific software development because future quantum software methods and abstractions must be usable not only by human experts, but also by autonomous coding agents that discover, compose, and optimize them through textual documentation and local experimentation.

Along the framework axis, TC demonstrates the highest overall task completion rate and performance efficiency among the evaluated frameworks. With the agent held fixed, TC successfully completes 10 out of 12 tasks, PennyLane reaches 8 of 12, and TorchQuantum and MindQuantum each reach 4 of 12 (Fig.~\ref{fig:dual}a). The artifact-runtime comparison gives a second separation: among passed tasks, TC has the smallest geometric mean slowdown across tasks relative to the expert TC reference, whereas the other tested frameworks show substantially larger slowdowns and thus longer artifact runtimes. This pattern should not be read only as an agent failing to find correct code in other frameworks. It also indicates that the tested frameworks expose different levels of native support and performance for these workloads, so framework capability and agent discoverability jointly determine the observed result.

Along the agent axis, current frontier systems show clear separation, though none achieve the performance of the human expert reference. On TC, Codex with GPT-5.5 reaches 10 of 12 passes, Claude Code with Claude Opus-4.8 reaches 9 of 12, Claude Code with Claude Sonnet-4.6 reaches 7 of 12, and Claude Code with GLM-5.2 reaches 6 of 12 (Fig.~\ref{fig:dual}b). The same configurations also differ strongly in artifact runtime on the tasks they solve. By construction, the expert TC reference is situated at zero failure and unit relative runtime. All evaluated configurations fall short of this baseline, demonstrating that state-of-the-art models do not yet reproduce the optimization quality of expert human developers.

\subsection*{Expert references reveal a remaining programming gap}

The expert TC reference implementations are essential for interpreting agent success. A boolean success metric can obscure the substantial gap between agent-generated and expert-level code. Among the tasks solved by agents, many implementations are much slower than the reference, and some numerically passing implementations fail semantic audit because they implement surrogate workflows inconsistent with the intended framework-native computation. The runtime gap is measured on the submitted solution itself, after the agent has finished, and therefore captures the computational quality of the generated research program rather than the cost of producing it. We report artifact runtime mainly as a ratio to the expert reference for the same task. This relative design is more portable than absolute wall time because it reduces dependence on the particular hardware used for a benchmark run.

\begin{figure*}[!t]
\centering
\includegraphics[width=\textwidth]{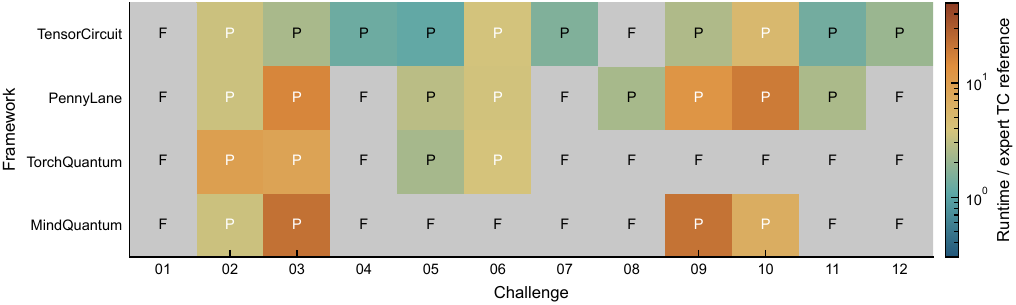}
\caption{\textbf{Task-level discrimination across quantum software frameworks.} Task-by-framework map for the fixed-solver framework comparison (Codex with GPT-5.5 model). Color encodes solution runtime relative to the expert TC reference for passed tasks where a timed solution is available; failed tasks are marked F. }
\label{fig:taskmap}
\end{figure*}

The distinction between agent-generated and expert-optimized program is visible in the runtime-ratio panels of Fig.~\ref{fig:dual} and in the task-level map of Fig.~\ref{fig:taskmap}. The expert references are not merely target solutions; they establish a practical baseline representing the performance ceiling achievable by expert developers with domain-specific knowledge of quantum algorithms and framework optimization. While current agents can generate functional solutions, they struggle to select optimal representations and algorithms, avoid costly graph staging or tensor contraction paths, and maintain the integrity of the physical process. Consequently, \orbit\ measures both functional success and the engineering gap between autonomous code generation and expert scientific software.

This expert-reference structure also makes the dataset valuable as a conventional software-performance benchmark. If the agent layer is removed, the same tasks define a suite of end-to-end scientific workloads on which quantum framework developers can compete to implement the fastest correct solution. Such a leaderboard would measure not only isolated kernel speed but also complete workflow performance across different quantum software frameworks.

\subsection*{The benchmark is discriminative rather than saturated}

An effective benchmark targeting state-of-the-art systems must avoid two primary failure modes: saturation, where high-performing systems cluster near the maximum score, and collapse, where most configurations fail uniformly. \orbit\ occupies a well-calibrated intermediate regime. Across frameworks with the same agent, success ranges from 4 of 12 to 10 of 12. Across agents on TC, success ranges from 6 of 12 to 10 of 12. Artifact efficiency also spans a wide range: the geometric mean slowdown on passed tasks ranges from about 2.2 times the expert reference for the best TC Codex configurations to about 10.1 times for the MindQuantum Codex configuration, with individual passed tasks extending from near-reference speed to more than 40 times slower than the expert baseline. The task-level map in Fig.~\ref{fig:taskmap} shows that failures are not concentrated in a single systemic bottleneck. Different frameworks and agents fail for different tasks and different reasons. 

This performance spread demonstrates that the suite has sufficient discriminative resolution to evaluate current and next-generation systems. The strongest configurations still exhibit failures and expert-reference gaps, leaving substantial headroom for development.

\subsection*{Failure patterns expose product-level and scientific limitations}

An analysis of failed evaluations reveals several distinct limits in agent-based scientific programming.
Some failures are standard software engineering failures, such as a missing solution artifact, timeout, or an implementation that does not satisfy the numerical evaluator.
Others are scientific-validity failures: the code can pass visible checks while bypassing the required framework, optimizing a surrogate objective, or omitting a required physical component.
A third class of failures arises from the guardrails of the deployed agent.
During the evaluation of the Claude-Opus-4.8 configuration, two tasks failed to complete because of cybersecurity refusals during local TC framework exploration, even though the benchmark task involved no network access, external service use, or cybersecurity objective. We classify these cases as false-positive safety refusals for this benchmark context and count them as failures because the deployed coding product did not deliver a valid research-software artifact.
This pattern is important for end-to-end evaluation: a model can be strong at code synthesis and still suffer reliability degradation through safety-layer misclassification, orchestration behavior, or tool-use policy decisions.

\begin{figure*}[!t]
\centering
\includegraphics[width=\textwidth]{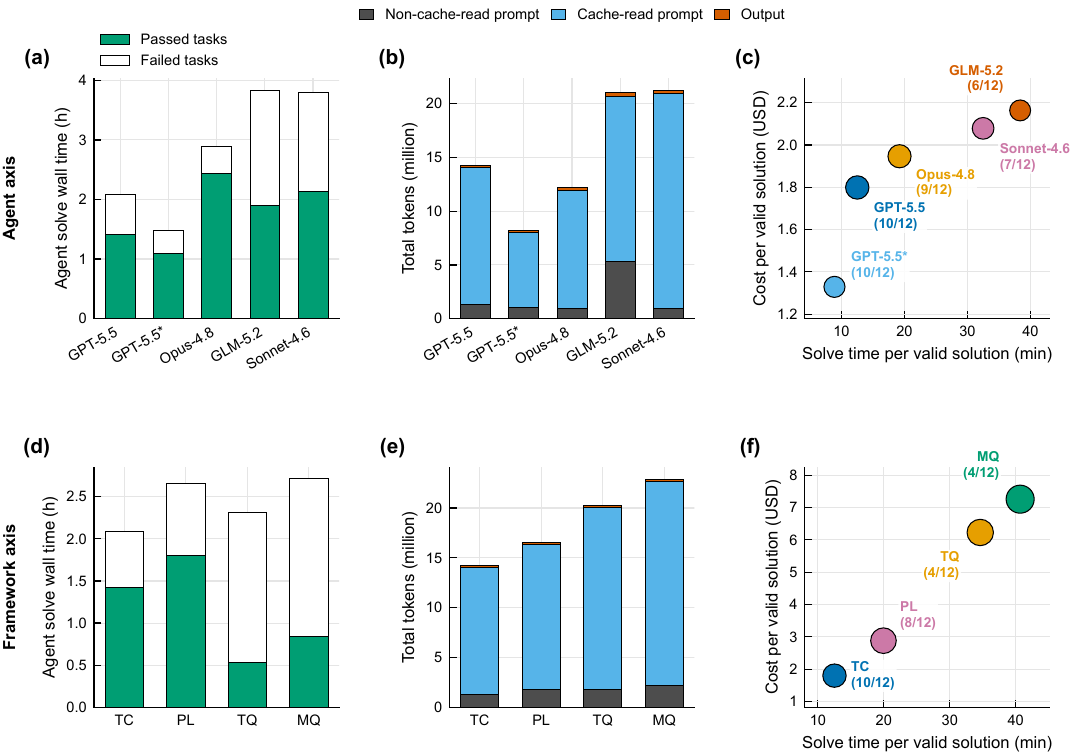}
\caption{\textbf{Agent-side resource use on both benchmark axes.} \textbf{a--c}, Model-configuration comparison on TC. \textbf{d--f}, Framework comparison for a fixed Codex agent using GPT-5.5. Panels \textbf{a} and \textbf{d} show agent solve wall time split by tasks that passed or failed; failed-task time is shown in white. Panels \textbf{b} and \textbf{e} show total solving-side token use split into non-cache-read prompt, cache-read prompt, and output tokens. Panels \textbf{c} and \textbf{f} show cost per valid solution versus solve time per valid solution; marker area scales with total solving-side cost. GPT-5.5* denotes GPT-5.5 with the TC-specific performance-checklist prompt. Abbreviations: TC, TensorCircuit-NG; PL, PennyLane; TQ, TorchQuantum; MQ, MindQuantum.}
\label{fig:consumption}
\end{figure*}

\subsection*{Agent-side resource utilization and economic efficiency}

Beyond task completion and artifact performance, \orbit\ evaluates the computational and financial resources consumed during the autonomous coding process, tracking agent solve wall time, token usage, and financial costs as shown in Fig.~\ref{fig:consumption}. Along the framework axis under a fixed agent (Codex with GPT-5.5), TC demonstrates the highest agent development efficiency, requiring the lowest total token overhead and shortest agentic solve duration compared to PennyLane, TorchQuantum, and MindQuantum (Fig.~\ref{fig:consumption}d-f).

Along the agent axis on TC, resource consumption varies significantly across frontier model configurations (Fig.~\ref{fig:consumption}a-c). Crucially, our systematic evaluations demonstrate that nominal token pricing is a poor predictor of the actual economic cost of scientific code generation. Models with lower unit prices per million tokens often incur higher cumulative expenses because their higher failure rates and need for extended exploratory tool-use drive up total token consumption. 

\subsection*{A performance checklist improves agent-side efficiency but not artifact-side efficiency}

We also tested a prompt-side intervention: a TC-specific performance checklist that provided generic TensorCircuit optimization hypotheses, including native TC representations, vectorization, JIT placement and reuse, static-data hoisting, sparse/matrix-product-operator/matrix-free observable paths, sampling choices, dtype and host-device-transfer discipline, and contraction-path-search choices. This checklist did not alter the functional success rate or overall artifact-level efficiency of the Codex GPT-5.5 TC configuration. However, it reduced agentic exploration and optimization overhead as measured by token use, estimated model-service cost, and solve duration over the evaluated tasks (Fig.~\ref{fig:consumption}a--c). The total token count decreased from about 14.24 million to about 8.21 million, and the solving-side estimated cost was reduced from US\$18.00 to US\$13.29.

This result distinguishes correctness-directed scaffolding from efficiency-directed prompting interventions. The checklist did not enable the agent to resolve the two remaining unsolved tasks under our evaluation criteria. Instead, it reduced the amount of agent-side exploration needed to reach implementation paths for tasks it was already capable of solving. The average artifact-level efficiency remained nearly unchanged: the geometric mean slowdown on passed tasks was about 2.2 with or without the checklist, although task-by-task runtimes shifted in both directions. For autonomous scientific programming systems, such interventions remain valuable because they optimize solve duration, token usage, and model-service expenses, but they should not be interpreted as closing the remaining expert-level artifact-performance gap.

\section*{Discussion}

\orbit\ reframes agentic code generation for scientific computing as an agent-framework co-performance problem. Rather than evaluating models in isolation, this dual-axis view evaluates the synergy between autonomous agents and scientific software ecosystems. This agent-native perspective is critical: it reveals which software abstractions are not only powerful for experts but also discoverable by agents, providing actionable guidance for the future design of scientific programming interfaces.

Our experimental findings yield three main conclusions. First, TC provides the most robust and compatible framework for agent-generated implementations in this domain, combining broad functionality with high performance. Second, a substantial gap remains between current frontier agents and human experts; while agents can solve many tasks, their artifact-side efficiency often lags significantly behind expert baselines. Third, benchmark economics must prioritize the cost per valid solution rather than nominal token pricing. Our analysis reveals a counterintuitive cost inversion: cheaper models often generate higher cumulative expenses due to excessive exploratory generation and failure-induced retries.

While \orbit\ establishes a rigorous foundation for evaluating scientific coding agents, several limitations present opportunities for future work. The semantic audit combines deterministic static checks, GPT-5.5-based source review, and human adjudication for borderline cases; future versions could further strengthen this layer with richer runtime provenance tracking and less human intervention. The framework comparison measures agent-mediated framework usability and performance rather than the best possible expert-written implementation for every framework. We nevertheless expect the artifact efficiency of agent-generated solutions to provide a meaningful proxy for frameworks' faithful performance, because agents can usually exploit the performant routes that the framework exposes. A complementary expert-optimized leaderboard across frameworks would be valuable for isolating framework capability and end-to-end performance without the agent-usability factor. Finally, while the compact task suite successfully provides discriminative resolution for current systems at a manageable cost, it remains a finite sample. Expanding this suite to encompass a wider spectrum of quantum programming paradigms is a critical next step.

Looking forward, \orbit\ can be extended with more frameworks, broader task families, public leaderboards, stronger provenance instrumentation, and expert baselines beyond TC. One natural leaderboard would rank agent-framework configurations by valid solutions, agent-side resource use, and artifact-level runtime. A second leaderboard would invite expert or team-optimized implementations in each framework and rank end-to-end scientific-computing performance on the same tasks. Ultimately, while \orbit\ uses quantum computing as its exemplar, its core principles establish a generalizable paradigm for evaluating autonomous code generation across all computational sciences.

\section*{Methods}

\subsection*{Task generation and framework constraints}

The benchmark tasks are derived from realistic quantum-research implementation scenarios. Each task exposes a problem statement and verifier but does not hardcode a quantum framework. Framework constraints are appended at run time through framework-specific prompt files. The selected framework is also passed to the execution environment and verifier through environment variables and image selection. This keeps the tasks framework-neutral while allowing strict framework-constrained evaluation. Additional implementation details, resource configuration, and prompt templates are provided in the Supplementary Information.

The maintained framework images share a common Dockerfile and install framework-specific Python requirements. The tested frameworks in this study are TC, PennyLane, TorchQuantum, and MindQuantum. The solver and verifier use the same framework image for a given evaluation, but the verifier is executed separately from the agent solution stage.

Framework inclusion requires that the framework expose a plausible native path for the differentiable workflows used by the benchmark. We therefore did not include Qiskit~\cite{javadiabhari2024qiskit} or Cirq~\cite{cirq2023} in the main evaluation. Although both are widely used circuit toolkits, their standard workflows are not designed around the end-to-end automatic-differentiation requirements of this task suite. Under the framework-native policy used here, writing a separate differentiable simulator around the toolkit would bypass the target framework rather than address the missing native support; such configurations would therefore be expected to fail most tasks rather than provide an informative comparison.

\subsection*{Solver protocol}

Each agent is required to create a Python solution exposing a function interface. The submitted solution must be concise and executable by the verifier. The solver interface supports command-line-based agents, including Codex and Claude Code, while maintaining a uniform execution boundary and verification pipeline. Solver-side wall time, token usage, and cost estimates are taken from the systematic evaluation logs.

\subsection*{Verifier protocol}

The verifier first runs the functional evaluator for the corresponding problem and measures the timed execution of the submitted solution. Static policy checks enforce line-count limits, required framework imports, forbidden framework imports, and obvious test or reward manipulation. The language-model semantic audit analyzes the source code to verify framework compliance and detect simulator bypasses,  or problem mismatches.

The final validity metric reported in this work combines functional correctness, static-policy validity, semantic-audit validity, and final human review for ambiguous framework-fidelity cases.

\subsection*{Artifact runtime, token and cost aggregation}

The whole agent evaluation runtime is based on Harbor framework. For task-level artifact-runtime comparisons, the evaluator-measured runtime of each submitted solution call is divided by the expert TC reference runtime for the same task. Geometric means are computed over passed tasks with available timed runtimes (900s). These artifact runtimes are distinct from solver wall time, which measures how long the agent took to produce the code. Agent costs are solving-side estimates computed from token accounting in the systematic evaluation logs. For Codex configurations, the estimate uses uncached input, cached input, and output token columns. For Claude Code configurations, the estimate uses uncached input, cache-write input, cache-read input, and output token columns. Verifier-side audit model usage is not included in the cost panels.

\subsection*{Performance-checklist prompt}

The performance checklist is an extra specific instruction only used for GPT-5.5* agent configuration. It does not provide task-specific answers. Instead, it gives generic workflow guidance for using framework-native implementation paths and testing performance choices before committing to a slow design. The checklist is evaluated as a workflow intervention by comparing success count, solve wall time, token use, solving-side cost, and artifact runtime against the baseline TC Codex GPT-5.5 configuration. The full checklist content is summarized in Supplementary Note 9.

\section*{Data availability}

The benchmark tasks, execution setup, verifier templates, and evaluation summaries are available in the accompanying repository: \url{https://github.com/sxzgroup/ORBIT-Q} and Supplementary Note 11.

\section*{Code availability}

The benchmark framework is available in the accompanying repository: \url{https://github.com/sxzgroup/ORBIT-Q}.

\section*{Acknowledgements}

This work was supported by the National Natural Science Foundation of China (No. 12504599 and No. 12574546), Quantum Science and Technology-National Science and Technology Major Project (No. 2024ZD0301700 and No. 2025ZD0300802), Science Challenge Project (No. TZ2025017), and the Chinese Academy of Sciences (No. XDB1680201 and No. YSBR-150).

\section*{Author contributions}

 S.-X.Z. and Y.-Q.C. conceived the benchmark direction, developed the benchmark infrastructure, ran the systematic evaluations, analyzed the results, and prepared the manuscript.

\section*{Competing interests}

S.-X.Z. and Y.-Q. C. are the authors of TensorCircuit-NG framework, which is evaluated in this study. The authors declare no other competing interests.

\bibliographystyle{apsreve}
\bibliography{ref}

@article{preskill2018nisq,
  author = {Preskill, John},
  title = {Quantum Computing in the {NISQ} era and beyond},
  journal = {Quantum},
  volume = {2},
  pages = {79},
  year = {2018},
  doi = {10.22331/q-2018-08-06-79}
}

@article{bharti2022nisq,
  author = {Bharti, Kishor and Cervera-Lierta, Alba and Kyaw, Thi Ha and Haug, Tobias and Alperin-Lea, Sumner and Anand, Abhinav and Degroote, Matthias and Heimonen, Hannu and Kottmann, Jakob S. and Menke, Tim and Mok, Wai-Keong and Sim, Sukin and Kwek, Leong-Chuan and Aspuru-Guzik, Al{\'a}n},
  title = {Noisy intermediate-scale quantum algorithms},
  journal = {Reviews of Modern Physics},
  volume = {94},
  pages = {015004},
  year = {2022},
  doi = {10.1103/RevModPhys.94.015004}
}

@article{cerezo2021vqa,
  author = {Cerezo, M. and Arrasmith, A. and Babbush, R. and Benjamin, S. C. and Endo, S. and Fujii, K. and McClean, J. R. and Mitarai, K. and Yuan, X. and Cincio, L. and Coles, P. J.},
  title = {Variational quantum algorithms},
  journal = {Nature Reviews Physics},
  volume = {3},
  pages = {625--644},
  year = {2021},
  doi = {10.1038/s42254-021-00348-9}
}

@article{biamonte2017qml,
  author = {Biamonte, Jacob and Wittek, Peter and Pancotti, Nicola and Rebentrost, Patrick and Wiebe, Nathan and Lloyd, Seth},
  title = {Quantum machine learning},
  journal = {Nature},
  volume = {549},
  pages = {195--202},
  year = {2017},
  doi = {10.1038/nature23474}
}

@article{tensorcircuit2023,
  author = {Zhang, Shi-Xin and Allcock, Jonathan and Wan, Zhou-Quan and Liu, Shuo and Sun, Jiace and Yu, Hao and Yang, Xing-Han and Qiu, Jiezhong and Ye, Zhaofeng and Chen, Yu-Qin and Lee, Chee-Kong and Zheng, Yi-Cong and Jian, Shao-Kai and Yao, Hong and Hsieh, Chang-Yu and Zhang, Shengyu},
  title = {{TensorCircuit}: a quantum software framework for the {NISQ} era},
  journal = {Quantum},
  volume = {7},
  pages = {912},
  year = {2023},
  doi = {10.22331/q-2023-02-02-912},
  eprint = {2205.10091},
  archivePrefix = {arXiv},
  primaryClass = {quant-ph}
}

@article{tensorcircuitng2026,
  author = {Zhang, Shi-Xin and Chen, Yu-Qin and Li, Weitang and Sun, Jiace and Ma, Wei-Guo and Zheng, Pei-Lin and Huang, Yu-Xiang and Wang, Qi-Xiang and Yu, Hui and Li, Zhuo and Huang, Xuyang and Li, Zong-Liang and Wan, Zhou-Quan and Liu, Shuo and Qiu, Jiezhong and Miao, Jiaqi and Song, Zixuan and Yan, Yuxuan and Tsuoka, Kazuki and Zhang, Pan and Wang, Lei and Fan, Heng and Hsieh, Chang-Yu and Yao, Hong and Xiang, Tao},
  title = {{TensorCircuit-NG}: A Universal, Composable, and Scalable Platform for Quantum Computing and Quantum Simulation},
  journal = {arXiv preprint arXiv:2602.14167},
  year = {2026},
  eprint = {2602.14167},
  archivePrefix = {arXiv},
  primaryClass = {quant-ph}
}

@article{pennylane2018,
  author = {Bergholm, Ville and Izaac, Josh and Schuld, Maria and Gogolin, Christian and Ahmed, Shahnawaz and Ajith, Vishnu and Alam, M. Sohaib and Alonso-Linaje, Guillermo and AkashNarayanan, B. and Asadi, Ali and Arrazola, Juan Miguel and Azad, Utkarsh and Banning, Sam and Blank, Carsten and Bromley, Thomas R. and Cordier, Benjamin A. and Ceroni, Jack and Delgado, Alain and Di Matteo, Olivia and Dusko, Amintor and Garg, Tanya and Guala, Diego and Hayes, Anthony and Hill, Ryan and Ijaz, Aroosa and Isacsson, Theodor and Ittah, David and Jahangiri, Soran and Jain, Prateek and Jiang, Edward and Khandelwal, Ankit and Kottmann, Korbinian and Lang, Robert A. and Lee, Christina and Loke, Thomas and Lowe, Angus and McKiernan, Keri and Meyer, Johannes Jakob and Monta{\~n}ez-Barrera, J. A. and Moyard, Romain and Niu, Zeyue and O'Riordan, Lee James and Oud, Steven and Panigrahi, Ashish and Park, Chae-Yeun and Polatajko, Daniel and Quesada, Nicol{\'a}s and Roberts, Chase and S{\'a}, Nahum and Schoch, Isidor and Shi, Borun and Shu, Shuli and Sim, Sukin and Singh, Arshpreet and Strandberg, Ingrid and Soni, Jay and Sz{\'a}va, Antal and Thabet, Slimane and Vargas-Hern{\'a}ndez, Rodrigo A. and Vincent, Trevor and Vitucci, Nicola and Weber, Maurice and Wierichs, David and Wiersema, Roeland and Willmann, Moritz and Wong, Vincent and Zhang, Shaoming and Killoran, Nathan},
  title = {{PennyLane}: Automatic differentiation of hybrid quantum-classical computations},
  journal = {arXiv preprint arXiv:1811.04968},
  year = {2018},
  eprint = {1811.04968},
  archivePrefix = {arXiv},
  primaryClass = {quant-ph}
}

@misc{torchquantum,
  author = {{MIT Han Lab}},
  title = {{TorchQuantum}: A {PyTorch}-based framework for quantum simulation and quantum machine learning},
  year = {2022},
  url = {https://github.com/mit-han-lab/torchquantum},
  note = {Software repository}
}

@article{mindsporequantum2024,
  author = {Xu, Xusheng and Cui, Jiangyu and Cui, Zidong and He, Runhong and Li, Qingyu and Li, Xiaowei and Lin, Yanling and Liu, Jiale and Liu, Wuxin and Lu, Jiale and Luo, Maolin and Lyu, Chufan and Pan, Shijie and Pavel, Mosharev and Shu, Runqiu and Tang, Jialiang and Xu, Ruoqian and Xu, Shu and Yang, Kang and Yu, Fan and Zeng, Qingguo and Zhao, Haiying and Zheng, Qiang and Zhou, Junyuan and Zhou, Xu and Zhu, Yikang and Zou, Zuoheng and Bayat, Abolfazl and Cao, Xi and Cui, Wei and Li, Zhendong and Long, Guilu and Su, Zhaofeng and Wang, Xiaoting and Wang, Zizhu and Wei, Shijie and Wu, Re-Bing and Zhang, Pan and Yung, Man-Hong},
  title = {{MindSpore Quantum}: A user-friendly, high-performance, and {AI}-compatible quantum computing framework},
  journal = {arXiv preprint arXiv:2406.17248},
  year = {2024},
  eprint = {2406.17248},
  archivePrefix = {arXiv},
  primaryClass = {quant-ph}
}

@article{openai2023gpt4,
  author = {{OpenAI}},
  title = {{GPT}-4 Technical Report},
  journal = {arXiv preprint arXiv:2303.08774},
  year = {2023},
  eprint = {2303.08774},
  archivePrefix = {arXiv},
  primaryClass = {cs.CL}
}

@article{deepseek2025r1,
  author = {{DeepSeek-AI}},
  title = {{DeepSeek-R1} Incentivizes Reasoning in {LLMs} Through Reinforcement Learning},
  journal = {Nature},
  volume = {645},
  pages = {633--638},
  year = {2025},
  doi = {10.1038/s41586-025-09422-z}
}

@article{chen2021codex,
  author = {Chen, Mark and Tworek, Jerry and Jun, Heewoo and Yuan, Qiming and Pinto, Henrique Ponde de Oliveira and Kaplan, Jared and Edwards, Harri and Burda, Yuri and Joseph, Nicholas and Brockman, Greg and Ray, Alex and Puri, Raul and Krueger, Gretchen and Petrov, Michael and Khlaaf, Heidy and Sastry, Girish and Mishkin, Pamela and Chan, Brooke and Gray, Scott and Ryder, Nick and Pavlov, Mikhail and Power, Alethea and Kaiser, Lukasz and Bavarian, Mohammad and Winter, Clemens and Tillet, Philippe and Such, Felipe Petroski and Cummings, Dave and Plappert, Matthias and Chantzis, Fotios and Barnes, Elizabeth and Herbert-Voss, Ariel and Guss, William Hebgen and Nichol, Alex and Paino, Alex and Tezak, Nikolas and Tang, Jie and Babuschkin, Igor and Balaji, Suchir and Jain, Shantanu and Saunders, William and Hesse, Christopher and Carr, Andrew N. and Leike, Jan and Achiam, Josh and Misra, Vedant and Morikawa, Evan and Radford, Alec and Knight, Matthew and Brundage, Miles and Murati, Mira and Mayer, Katie and Welinder, Peter and McGrew, Bob and Amodei, Dario and McCandlish, Sam and Sutskever, Ilya and Zaremba, Wojciech},
  title = {Evaluating large language models trained on code},
  journal = {arXiv preprint arXiv:2107.03374},
  year = {2021},
  eprint = {2107.03374},
  archivePrefix = {arXiv},
  primaryClass = {cs.LG}
}

@inproceedings{jimenez2024swebench,
  author = {Jimenez, Carlos E. and Yang, John and Wettig, Alexander and Yao, Shunyu and Pei, Kexin and Press, Ofir and Narasimhan, Karthik R.},
  title = {{SWE}-bench: Can language models resolve real-world {GitHub} issues?},
  booktitle = {International Conference on Learning Representations},
  year = {2024},
  url = {https://arxiv.org/abs/2310.06770},
  eprint = {2310.06770},
  archivePrefix = {arXiv},
  primaryClass = {cs.CL}
}

@inproceedings{yang2024sweagent,
  author = {Yang, John and Jimenez, Carlos E. and Wettig, Alexander and Lieret, Kilian and Yao, Shunyu and Narasimhan, Karthik R. and Press, Ofir},
  title = {{SWE}-agent: Agent-computer interfaces enable automated software engineering},
  booktitle = {Advances in Neural Information Processing Systems 37},
  pages = {50528--50652},
  year = {2024},
  doi = {10.52202/079017-1601},
  url = {https://arxiv.org/abs/2405.15793},
  eprint = {2405.15793},
  archivePrefix = {arXiv},
  primaryClass = {cs.SE}
}

@article{austin2021mbpp,
  author = {Austin, Jacob and Odena, Augustus and Nye, Maxwell and Bosma, Maarten and Michalewski, Henryk and Dohan, David and Jiang, Ellen and Cai, Carrie and Terry, Michael and Le, Quoc and Sutton, Charles},
  title = {Program Synthesis with Large Language Models},
  journal = {arXiv preprint arXiv:2108.07732},
  year = {2021},
  eprint = {2108.07732},
  archivePrefix = {arXiv},
  primaryClass = {cs.PL}
}

@inproceedings{hendrycks2021apps,
  author = {Hendrycks, Dan and Basart, Steven and Kadavath, Saurav and Mazeika, Mantas and Arora, Akul and Guo, Ethan and Burns, Collin and Puranik, Samir and He, Horace and Song, Dawn and Steinhardt, Jacob},
  title = {Measuring Coding Challenge Competence With {APPS}},
  booktitle = {Proceedings of the Neural Information Processing Systems Track on Datasets and Benchmarks},
  volume = {1},
  year = {2021},
  url = {https://arxiv.org/abs/2105.09938},
  eprint = {2105.09938},
  archivePrefix = {arXiv},
  primaryClass = {cs.SE}
}

@inproceedings{jain2024livecodebench,
  author = {Jain, Naman and Han, King and Gu, Alex and Li, Wen-Ding and Yan, Fanjia and Zhang, Tianjun and Wang, Sida and Solar-Lezama, Armando and Sen, Koushik and Stoica, Ion},
  title = {{LiveCodeBench}: Holistic and Contamination Free Evaluation of Large Language Models for Code},
  booktitle = {International Conference on Learning Representations},
  year = {2025},
  url = {https://arxiv.org/abs/2403.07974},
  eprint = {2403.07974},
  archivePrefix = {arXiv},
  primaryClass = {cs.SE}
}

@article{steiger2018projectq,
  author = {Steiger, Damian S. and H{\"a}ner, Thomas and Troyer, Matthias},
  title = {{ProjectQ}: An Open Source Software Framework for Quantum Computing},
  journal = {Quantum},
  volume = {2},
  pages = {49},
  year = {2018},
  doi = {10.22331/q-2018-01-31-49}
}

@article{javadiabhari2024qiskit,
  author = {Javadi-Abhari, Ali and Treinish, Matthew and Krsulich, Kevin and Wood, Christopher J. and Lishman, Jake and Gacon, Julien and Martiel, Simon and Nation, Paul D. and Bishop, Lev S. and Cross, Andrew W. and Johnson, Blake R. and Gambetta, Jay M.},
  title = {Quantum Computing with {Qiskit}},
  journal = {arXiv preprint arXiv:2405.08810},
  year = {2024},
  eprint = {2405.08810},
  archivePrefix = {arXiv},
  primaryClass = {quant-ph}
}

@article{broughton2021tfq,
  author = {Broughton, Michael and Verdon, Guillaume and McCourt, Trevor and Martinez, Antonio J. and Yoo, Jae Hyeon and Isakov, Sergei V. and Massey, Philip and Halavati, Ramin and Niu, Murphy Yuezhen and Zlokapa, Alexander and Peters, Evan and Lockwood, Owen and Skolik, Andrea and Jerbi, Sofiene and Dunjko, Vedran and Leib, Martin and Streif, Michael and Von Dollen, David and Chen, Hongxiang and Cao, Shuxiang and Wiersema, Roeland and Huang, Hsin-Yuan and McClean, Jarrod R. and Babbush, Ryan and Boixo, Sergio and Bacon, Dave and Ho, Alan K. and Neven, Hartmut and Mohseni, Masoud},
  title = {{TensorFlow Quantum}: A Software Framework for Quantum Machine Learning},
  journal = {arXiv preprint arXiv:2003.02989},
  year = {2020},
  eprint = {2003.02989},
  archivePrefix = {arXiv},
  primaryClass = {quant-ph}
}

@article{johansson2012qutip,
  author = {Johansson, J. R. and Nation, P. D. and Nori, Franco},
  title = {{QuTiP}: An Open-source {Python} Framework for the Dynamics of Open Quantum Systems},
  journal = {Computer Physics Communications},
  volume = {183},
  number = {8},
  pages = {1760--1772},
  year = {2012},
  doi = {10.1016/j.cpc.2012.02.021}
}

@article{gray2018quimb,
  author = {Gray, Johnnie},
  title = {{quimb}: A {Python} Package for Quantum Information and Many-body Calculations},
  journal = {Journal of Open Source Software},
  volume = {3},
  number = {29},
  pages = {819},
  year = {2018},
  doi = {10.21105/joss.00819}
}

@article{gray2021cotengra,
  author = {Gray, Johnnie and Kourtis, Stefanos},
  title = {Hyper-optimized Tensor Network Contraction},
  journal = {Quantum},
  volume = {5},
  pages = {410},
  year = {2021},
  doi = {10.22331/q-2021-03-15-410}
}

@article{peruzzo2014vqe,
  author = {Peruzzo, Alberto and McClean, Jarrod and Shadbolt, Peter and Yung, Man-Hong and Zhou, Xiao-Qi and Love, Peter J. and Aspuru-Guzik, Al{\'a}n and O'Brien, Jeremy L.},
  title = {A Variational Eigenvalue Solver on a Photonic Quantum Processor},
  journal = {Nature Communications},
  volume = {5},
  pages = {4213},
  year = {2014},
  doi = {10.1038/ncomms5213}
}

@article{mcclean2016vqa,
  author = {McClean, Jarrod R. and Romero, Jonathan and Babbush, Ryan and Aspuru-Guzik, Al{\'a}n},
  title = {The Theory of Variational Hybrid Quantum-Classical Algorithms},
  journal = {New Journal of Physics},
  volume = {18},
  pages = {023023},
  year = {2016},
  doi = {10.1088/1367-2630/18/2/023023}
}

@article{farhi2014qaoa,
  author = {Farhi, Edward and Goldstone, Jeffrey and Gutmann, Sam},
  title = {A Quantum Approximate Optimization Algorithm},
  journal = {arXiv preprint arXiv:1411.4028},
  year = {2014},
  eprint = {1411.4028},
  archivePrefix = {arXiv},
  primaryClass = {quant-ph}
}

@article{schuld2019gradients,
  author = {Schuld, Maria and Bergholm, Ville and Gogolin, Christian and Izaac, Josh and Killoran, Nathan},
  title = {Evaluating Analytic Gradients on Quantum Hardware},
  journal = {Physical Review A},
  volume = {99},
  pages = {032331},
  year = {2019},
  doi = {10.1103/PhysRevA.99.032331}
}

@article{yuan2021hybridtn,
  author = {Yuan, Xiao and Sun, Jinzhao and Liu, Junyu and Zhao, Qi and Zhou, You},
  title = {Quantum Simulation with Hybrid Tensor Networks},
  journal = {Physical Review Letters},
  volume = {127},
  pages = {040501},
  year = {2021},
  doi = {10.1103/PhysRevLett.127.040501}
}

@article{zhang2022vqnhe,
  author = {Zhang, Shi-Xin and Wan, Zhou-Quan and Lee, Chee-Kong and Hsieh, Chang-Yu and Zhang, Shengyu and Yao, Hong},
  title = {Variational Quantum-Neural Hybrid Eigensolver},
  journal = {Physical Review Letters},
  volume = {128},
  pages = {120502},
  year = {2022},
  doi = {10.1103/PhysRevLett.128.120502}
}

@article{chen2026qmlresilience,
  author = {Chen, Yu-Qin and Zhang, Shi-Xin},
  title = {Superior Resilience to Poisoning and Amenability to Unlearning in Quantum Machine Learning},
  journal = {Nature Communications},
  volume = {17},
  pages = {3716},
  year = {2026},
  doi = {10.1038/s41467-026-70420-4}
}

@article{orus2014tensor,
  author = {Orús, Román},
  title = {A Practical Introduction to Tensor Networks: Matrix Product States and Projected Entangled Pair States},
  journal = {Annals of Physics},
  volume = {349},
  pages = {117--158},
  year = {2014},
  doi = {10.1016/j.aop.2014.06.013}
}

@article{schollwock2011dmrg,
  author = {Schollwöck, Ulrich},
  title = {The Density-matrix Renormalization Group in the Age of Matrix Product States},
  journal = {Annals of Physics},
  volume = {326},
  number = {1},
  pages = {96--192},
  year = {2011},
  doi = {10.1016/j.aop.2010.09.012}
}

@article{white1992dmrg,
  author = {White, Steven R.},
  title = {Density Matrix Formulation for Quantum Renormalization Groups},
  journal = {Physical Review Letters},
  volume = {69},
  pages = {2863--2866},
  year = {1992},
  doi = {10.1103/PhysRevLett.69.2863}
}

@article{vidal2003simulation,
  author = {Vidal, Guifr{\'e}},
  title = {Efficient Classical Simulation of Slightly Entangled Quantum Computations},
  journal = {Physical Review Letters},
  volume = {91},
  pages = {147902},
  year = {2003},
  doi = {10.1103/PhysRevLett.91.147902}
}

@article{markov2008tensor,
  author = {Markov, Igor L. and Shi, Yaoyun},
  title = {Simulating Quantum Computation by Contracting Tensor Networks},
  journal = {SIAM Journal on Computing},
  volume = {38},
  number = {3},
  pages = {963--981},
  year = {2008},
  doi = {10.1137/050644756}
}

@article{cross2019quantumvolume,
  author = {Cross, Andrew W. and Bishop, Lev S. and Sheldon, Sarah and Nation, Paul D. and Gambetta, Jay M.},
  title = {Validating Quantum Computers Using Randomized Model Circuits},
  journal = {Physical Review A},
  volume = {100},
  pages = {032328},
  year = {2019},
  doi = {10.1103/PhysRevA.100.032328}
}

@inproceedings{tomesh2022supermarq,
  author = {Tomesh, Teague and Gokhale, Pranav and Omole, Victory and Ravi, Gokul Subramanian and Smith, Kaitlin N. and Viszlai, Joshua and Wu, Xin-Chuan and Hardavellas, Nikos and Martonosi, Margaret R. and Chong, Frederic T.},
  title = {{SupermarQ}: A Scalable Quantum Benchmark Suite},
  booktitle = {2022 IEEE International Symposium on High-Performance Computer Architecture (HPCA)},
  pages = {587--603},
  year = {2022},
  doi = {10.1109/HPCA53966.2022.00050}
}

@article{li2023qasmbench,
  author = {Li, Ang and Stein, Samuel and Krishnamoorthy, Sriram and Ang, James},
  title = {{QASMBench}: A Low-Level Quantum Benchmark Suite for {NISQ} Evaluation and Simulation},
  journal = {ACM Transactions on Quantum Computing},
  volume = {4},
  number = {2},
  pages = {1--26},
  year = {2023},
  doi = {10.1145/3550488}
}

@article{quetschlich2023mqtbench,
  author = {Quetschlich, Nils and Burgholzer, Lukas and Wille, Robert},
  title = {{MQT Bench}: Benchmarking Software and Design Automation Tools for Quantum Computing},
  journal = {Quantum},
  volume = {7},
  pages = {1062},
  year = {2023},
  doi = {10.22331/q-2023-07-20-1062}
}

@misc{cirq2023,
  author = {{Cirq Developers}},
  title = {Cirq},
  year = {2023},
  url = {https://github.com/quantumlib/Cirq},
  note = {Software repository}
}

@article{harris2020array,
  title = {Array programming with {NumPy}},
  author = {Harris, Charles R. and Millman, K. Jarrod and van der Walt, Stéfan J. and Gommers, Ralf and Virtanen, Pauli and Cournapeau, David and Wieser, Eric and Taylor, Julian and Berg, Sebastian and Smith, Nathaniel J. and Kern, Robert and Picus, Matti and Hoyer, Stephan and van Kerkwijk, Marten H. and Brett, Matthew and Haldane, Allan and del Río, Jaime Fernández and Wiebe, Mark and Peterson, Pearu and Gérard-Marchant, Pierre and Sheppard, Kevin and Reddy, Tyler and Weckesser, Warren and Abbasi, Hameer and Gohlke, Christoph and Oliphant, Travis E.},
  journal = {Nature},
  volume = {585},
  number = {7825},
  pages = {357--362},
  year = {2020},
  doi = {10.1038/s41586-020-2649-2}
}

@misc{jax2018github,
  author = {James Bradbury and Roy Frostig and Peter Hawkins and Matthew James Johnson and Chris Leary and Dougal Maclaurin and George Necula and Adam Paszke and Jake Vander{P}las and Skye Wanderman-{M}ilne and Qiao Zhang},
  title = {{JAX}: composable transformations of {Python}+{NumPy} programs},
  year = {2018},
  url = {https://github.com/google/jax},
  note = {Software repository}
}

@article{Chen2025IntrinsicPreservation,
    author = {Chen, Yu-Qin and Zhang, Shi-Xin},
    title = {Intrinsic Preservation of Plasticity in Continual Quantum Learning},
    journal = {arXiv preprint arXiv:2511.17228},
    year = {2025},
    eprint = {2511.17228},
    archivePrefix = {arXiv},
    primaryClass = {quant-ph}
  }

@article{Zhang2026QuantumSubliminalLearning,
    author = {Zhang, Shi-Xin and Chen, Yu-Qin},
    title = {Quantum Subliminal Learning},
    journal = {arXiv preprint arXiv:2605.29557},
    year = {2026},
    eprint = {2605.29557},
    archivePrefix = {arXiv},
    primaryClass = {quant-ph}
  }

@article{Hu2019QuantumGenerative,
    author = {Hu, Ling and Wu, Shu-Hao and Cai, Weizhou and Ma, Yuwei and Mu, Xianghao and Xu, Yuan and Wang, Haiyan and Song, Yipu and Deng, Dong-
    Ling and Zou, Chang-Ling and Sun, Luyan},
    title = {Quantum Generative Adversarial Learning in a Superconducting Quantum Circuit},
    journal = {Science Advances},
    volume = {5},
    number = {1},
    pages = {eaav2761},
    year = {2019},
    doi = {10.1126/sciadv.aav2761}
  }

@article{Zhou2020QuantumApproximate,
    author = {Zhou, Leo and Wang, Sheng-Tao and Choi, Soonwon and Pichler, Hannes and Lukin, Mikhail D.},
    title = {Quantum Approximate Optimization Algorithm: Performance, Mechanism, and Implementation on Near-Term Devices},
    journal = {Physical Review X},
    volume = {10},
    number = {2},
    pages = {021067},
    year = {2020},
    doi = {10.1103/PhysRevX.10.021067}
  }

@article{Cheng2024QuantumApproximate,
    author = {Cheng, Lixue and Chen, Yu-Qin and Zhang, Shi-Xin and Zhang, Shengyu},
    title = {Quantum Approximate Optimization via Learning-Based Adaptive Optimization},
    journal = {Communications Physics},
    volume = {7},
    number = {1},
    pages = {83},
    year = {2024},
    doi = {10.1038/s42005-024-01577-x}
  }

\end{document}


\maketitle

\section*{Supplementary Note 1: Framework-neutral benchmark construction}

\orbit\ separates the scientific task definition from the quantum-software framework constraint.
Each challenge provides a problem statement, a solution interface, and a verifier.
The task statement describes the physical workflow, the expected return values, and the functional evaluator, but it does not hardcode a particular quantum package.
The tasks were selected from realistic quantum-research implementation patterns and collectively cover a broad range of framework capabilities required in modern simulation and optimization workflows.
The selected framework enters each evaluation through three controlled channels: a framework-specific instruction block appended to the agent prompt, a framework-specific execution image, and verifier policy variables that define which framework is required for source-level compliance.
This design supports two orthogonal experimental axes.
In the agent axis, the framework is held fixed while the agent configuration is varied.
In the framework axis, the agent is held fixed while the required quantum framework is varied.
Because every implementation on both axes is produced by an autonomous agent under a unified harness, the framework comparison is decoupled from human manual optimization.
It instead measures the co-performance and synergy of framework coverage, expressivity, performance, application programming interface (API) discoverability, and agent usability.
The same task definitions can also be used without an agent layer.
In that setting, \orbit\ becomes a conventional end-to-end benchmark for quantum software frameworks, where expert-written implementations can be compared on feature coverage, differentiability support, and end-to-end runtime.
This use case is complementary to agentic evaluation because it fully isolates the software stack's best achievable implementation quality from the agent's ability to discover that implementation.

\section*{Supplementary Note 2: Framework inclusion and exclusion criteria}

Framework inclusion requires a plausible native route for the differentiable and research-grade workflows used by \orbit.
The main evaluation therefore focuses on differentiable quantum software frameworks for which the benchmark could be run under a comparable local containerized setup.
We excluded TensorFlow Quantum from the main evaluation because its public releases lack native support for macOS/ARM architectures, preventing compatibility with our Apple-silicon-based development and Docker evaluation pipeline.
More importantly, TensorFlow Quantum's high-level abstraction model is less directly aligned with several \orbit\ tasks that require fine-grained control over nonstandard simulation workflows.
Under the same framework-native validity policy used for the other systems, this would likely lead to weak task coverage and poor end-to-end performance rather than an informative competitive comparison.
Similarly, Qiskit and Cirq are important circuit-construction ecosystems, but their standard workflows are not designed around the end-to-end automatic-differentiation requirements of this task suite.
Because \orbit\ explicitly disallows framework bypass, agents could not compensate for missing native support by constructing a separate differentiable simulator around these toolkits.
These configurations would therefore be expected to fail most tasks under the present validity policy rather than provide an informative comparison.

\section*{Supplementary Note 3: Harbor adaptation for quantum-framework evaluation}

The benchmark uses Harbor as the orchestration layer but customizes the environment, agent, and verifier interfaces for framework-constrained quantum programming.
The canonical task remains framework-neutral.
At runtime, the wrapper selects the challenge, the required framework, the agent, the agent model, the audit model, the framework prompt, and the framework image.
The framework image is selected by a framework-aware environment adapter rather than by modifying the canonical task.
A framework-specific instruction block is appended to the task prompt and states both the required framework and the prohibited bypass modes.
The verifier receives the required framework as a policy variable and runs functional evaluation, static checks, and source-level semantic audit under the same framework environment used by the agent.
The agent interface invokes Codex or Claude Code command-line agents while preserving the same task interface and verifier interface.
Codex runs used \texttt{high} reasoning effort, whereas Claude Code runs used \texttt{max} reasoning effort.
The task-level interface is intentionally minimal.
Each agent must produce a Python source file exposing \texttt{run\_solution(config)} and must return NumPy-compatible outputs consumed by the evaluator.
This prevents framework-specific task rewrites while still allowing each framework to express the core quantum computation through its native primitives.

\section*{Supplementary Note 4: Resources and execution limits}

All evaluations were performed in a containerized environment with standardized resource allocations of 14 CPU cores, 32 GiB of unified memory, and 16 GiB of local storage. The evaluations were performed on a host system equipped with an Apple M4 Pro processor.
The agent time budget was 1800 s.
The verifier time budget was 1200 s.
The timing observable refers to artifact-side runtime: the elapsed time of the submitted solution call after the agent has already produced a source file.
The principal artifact-runtime statistic is the ratio between this submitted-solution runtime and the expert \tcng\ (TC) reference runtime for the same task.
This relative measure is more portable across hardware than reporting only absolute wall time.
This metric is distinct from agent-side wall time, token use, and monetary cost, which quantify the resources consumed while generating the source file.
The distinction is important because an agent can be inexpensive while producing a slow scientific program, or expensive while producing no valid framework-native artifact.
The framework images share a common base environment with Python 3.11, command-line coding-agent tools, build tools, and common scientific Python packages.
Framework-specific requirements are installed in separate images.
The same selected framework image is used by the agent and verifier for a given evaluation, which prevents discrepancies between the implementation environment and the evaluation environment.

\section*{Supplementary Note 5: Dataset use as a quantum-software performance benchmark}

The \orbit\ task suite is intended to support two related forms of evaluation.
The first is the agent-framework evaluation reported in the main text, where every submitted implementation is generated by an autonomous coding agent.
The second is a framework-performance evaluation in which expert developers implement the same tasks directly in each framework.
The second mode does not measure agent usability, but it provides a systematic benchmark for framework support and end-to-end runtime on research-grade quantum workloads.
The tasks are deliberately broader than isolated kernels or small-circuit demonstrations.
A framework that performs well across the full suite must therefore support broad research-grade programming pipelines rather than a single optimized primitive.
For this reason, \orbit\ can support public leaderboards that rank valid implementations by wall-clock artifact runtime and memory behavior under a shared evaluator.
Such leaderboards would provide actionable targets for quantum-software developers: optimizing contraction path search, reducing graph-compilation overhead, improving high-level primitives, stabilizing automatic differentiation, and exposing APIs that make performant routes discoverable.
The expert TC reference implementations used in the present study provide one reference point for this broader performance-benchmarking mode, but the dataset does not require future expert baselines to be limited to TC.

\section*{Supplementary Note 6: Validity criteria and source-level semantic audit}

The validity decision combines functional correctness, static policy checks, source-level semantic audit, and final human adjudication for ambiguous framework-fidelity cases.
Functional correctness checks whether the submitted result satisfies the evaluator for the corresponding scientific task.
Static policy checks include line-count limits, required framework imports, forbidden quantum-framework imports, and obvious manipulation of tests or reward files.
The source-level semantic audit is performed with GPT-5.5 throughout this work and is necessary because scientific quantum programming can fail even when surface-level numerical checks pass.
A solution is invalid if it uses the requested framework only as a gate-matrix helper while the core quantum evolution is performed by custom NumPy or JAX code.
A solution is invalid if it substitutes different physical objectives, synthesizes target histories or trajectories, replaces exact differentiable workflows with non-differentiable surrogates, or exploits evaluator threshold weaknesses.
The audit prompt requests a structured JSON object containing the following fields.
\begin{itemize}
\item The \codefield{uses_required_framework} field indicates whether the submitted source uses the required framework for the core quantum computation.
\item The \codefield{imports_other_quantum_framework} field indicates whether the source imports a non-required quantum framework.
\item The \codefield{uses_other_quantum_framework} field indicates whether a non-required quantum framework is used for the quantum computation.
\item The \codefield{raw_numpy_jax_quantum_simulator_bypass} field indicates whether custom array code bypasses the required framework.
\item The \codefield{hardcoded_or_hidden_answer} field indicates whether the solution appears to hardcode an oracle or hidden answer.
\item The \codefield{tampers_with_tests_or_rewards} field indicates whether the solution manipulates tests, logs, or reward files.
\item The \codefield{exploits_evaluator_weakness} field indicates whether the solution exploits evaluator-specific weaknesses rather than solving the stated task.
\item The \codefield{faithfully_implements_problem} field indicates whether the source follows the stated physical workflow.
\item The \codefield{obvious_implementation_error} field indicates whether the source contains an evident logic or implementation error.
\item The \codefield{problem_alignment_issues} field lists concrete discrepancies between the implementation and the problem statement.
\item The \codefield{framework_rationale} field explains the framework-compliance decision.
\item The \codefield{cheating_rationale} field explains the cheating and evaluator-exploit decision.
\item The \codefield{fidelity_rationale} field explains the problem-fidelity decision.
\item The \codefield{implementation_error_rationale} field explains the implementation-correctness decision.
\item The \codefield{confidence} field indicates the auditor's confidence in the structured decision.
\end{itemize}
The language-model semantic-audit score combines framework compliance, integrity compliance, problem fidelity, and absence of obvious implementation error.
The audit result is interpreted together with functional output and final human review for subtle cases.

\section*{Supplementary Note 7: Failure-pattern adjudication}

Failed evaluations were assigned one primary invalidity mode after inspecting the available agent artifact, verifier output, and structured run metadata.
Supplementary Table~\ref{tab:failure-mode-attribution} summarizes this attribution for the TC model-axis evaluations.
The agent/product category covers agent timeouts, tool or process exits, and missing solution artifacts before a scorable implementation is produced.
The resource timeout/out-of-memory (OOM) category covers submitted artifacts that enter verifier execution but time out or are killed by resource pressure.
The runtime/API-exception category covers submitted artifacts that parse successfully but crash from framework API or type errors during execution.
The framework-bypass category covers implementations where the required framework is used only peripherally while the core simulator, contraction, sampling, or optimization path is handwritten.
The problem-mismatch category covers surrogate physical workflows, omitted required initialization, manufactured histories, or using a state-preparation path inconsistent with the stated variational procedure.
The safety-refusal category covers benign scientific-programming tasks that terminate because the agent refuses to proceed.
No selected TC model-axis failure was attributed to a Python syntax error.

These labels describe the proximal invalidity mode of the submitted artifact rather than a complete causal explanation.
In cross-framework comparisons, trace inspection showed that some problem-mismatch or framework-bypass cases arose only after the agent first attempted a framework-native route but encountered missing, inaccessible, or poorly performant framework capabilities.
Such failures can therefore reflect the framework's effective task coverage and API affordances as much as the agent's scientific-programming ability.
By contrast, within the fixed TC model-axis evaluation, where validated reference routes exist for all tasks under the same framework constraint, the same attribution more directly probes differences in agent capability.

This attribution separates product-level completion failures from scientific or framework-fidelity failures in a submitted artifact.
The Opus-4.8 configuration illustrates the distinction.
One failed task produced a solution artifact but failed functional evaluation by timeout.
Two other failed tasks terminated before producing a solution artifact because the agent classified local TC API exploration as a cyber-safety case and refused to continue.
These two evaluations are counted as failures because the evaluated product did not complete the task, but they are not interpreted as evidence that the underlying quantum algorithms or TC primitives were intrinsically infeasible.
They instead show that end-to-end coding-agent benchmarks evaluate the integrated agent system, including safety classifiers and tool-use policies, rather than merely the base model's programming capabilities.

\begin{table}[htbp]
\centering
\scriptsize
\setlength{\tabcolsep}{2pt}
\caption{\textbf{Primary failure-mode attribution for TC model-axis evaluations.} Counts are over the 12 TensorCircuit-constrained tasks for each model configuration. GPT-5.5* denotes GPT-5.5 with the TC performance-checklist prompt.}
\label{tab:failure-mode-attribution}
\begin{tabular}{@{}lcccccccc@{}}
\toprule
 & \multicolumn{2}{c}{Outcome} & \multicolumn{6}{c}{Primary invalidity mode} \\
\cmidrule(lr){2-3}\cmidrule(l){4-9}
Model & Valid & Invalid & \shortstack{Agent/product\\no artifact} & \shortstack{Resource\\timeout/OOM} & \shortstack{Runtime/API\\exception} & \shortstack{Framework\\bypass} & \shortstack{Problem\\mismatch} & \shortstack{Safety\\refusal} \\
\midrule
GPT-5.5 & 10 & 2 & 1 & 0 & 0 & 1 & 0 & 0 \\
GPT-5.5* & 10 & 2 & 0 & 0 & 0 & 1 & 1 & 0 \\
Opus-4.8 & 9 & 3 & 0 & 1 & 0 & 0 & 0 & 2 \\
GLM-5.2 & 6 & 6 & 3 & 1 & 0 & 0 & 2 & 0 \\
Sonnet-4.6 & 7 & 5 & 2 & 1 & 1 & 1 & 0 & 0 \\
\bottomrule
\end{tabular}
\end{table}

\section*{Supplementary Note 8: Framework-constraint prompt template}

The framework-constraint prompt is appended to every task, with \texttt{\{framework\}} replaced by the selected framework name.
The full template is shown in Supplementary Listing~\ref{lst:framework-prompt}.

\begin{lstlisting}[style=promptlisting,caption={Framework-constraint prompt template.},label={lst:framework-prompt}]
## Quantum Framework Constraint

The required quantum software framework for this run is {framework}.

You may use general-purpose Python libraries for numerical arrays, optimization, sparse linear algebra, data handling, testing, and interoperability.

However:

- The central quantum computation must be expressed through {framework}.
- Do not use another quantum software framework to implement the task.
- Do not replace {framework} with a raw NumPy/JAX/SciPy simulator from scratch.
- Do not implement the required quantum gates, circuit evolution, measurements, or state updates by manually manipulating the full state vector, amplitudes, tensor entries, or index tables as a workaround around the framework. Use the quantum software framework's own circuit/state/operator APIs for the core quantum process.
- Do not use {framework} only as a gate-matrix lookup helper, state initializer, or thin wrapper while the real quantum evolution is carried out by custom array code. The implementation itself must genuinely run through {framework}'s quantum APIs.
- Do not compute an exact or near-exact oracle answer for the evaluator's target quantity and then use that oracle value to fabricate outputs that merely satisfy the verifier checks.
- Do not directly diagonalize, exactly solve, or otherwise compute the strict target optimum of the same Hamiltonian/objective when the task is asking for a variational, iterative, sampled, or framework-native optimization procedure, unless the task explicitly allows that method.
- Do not fake, smooth, interpolate, backfill, or analytically construct required return arrays such as `energy_history`, `loss_history`, gradients, entropies, measurement statistics, or trajectory records. Every returned entry must come from the actual required computation at the corresponding step.
- Do not exploit evaluator loopholes by returning values that pass surface-level checks while skipping the required optimization, update loop, circuit evolution, measurement process, or other requested algorithmic structure.
- Treat the problem description itself as part of the specification, not just the visible evaluator outputs. The code and implementation workflow themselves must be consistent with the described physics process, optimization procedure, observables, measurement or post-selection rules, and returned quantities.
- Assume there may be additional hidden or black-box evaluation derived from the problem statement, not only the visible checks in the provided evaluator script. A solution that only targets the visible checks while violating the stated physics or algorithmic process is invalid.
- Do not hardcode parameter targets, closed-form trajectories, fixed entropy profiles, synthetic measurement statistics, or any other surrogate outputs unless they are genuinely produced by the required computation described in the task.
- If the stated physical process says a quantity should come from circuit evolution, variational updates, measurements, post-selection, entropy evaluation, or trajectory averaging, then return the value produced by that process, not a proxy chosen merely to satisfy the visible thresholds.
- If the problem specifies Adam, gradient descent, stochastic gradient descent, natural gradient, coordinate descent, or any other named optimizer or update rule, then the code must actually implement that optimizer workflow on the stated objective. Do not substitute a different update rule, a hand-chosen target parameter vector, or a synthetic monotone trajectory.
- When the required optimization workflow depends on gradients, use exact or mathematically strict gradients for the stated objective. Automatic differentiation through the required framework is preferred. Parameter-shift or another exact analytic gradient method is also allowed when correct. Finite-difference gradients, SPSA, and other approximate-gradient substitutes are not allowed.
- If the problem specifies one value per optimizer update, per time step, per measurement event, per block, or per trajectory, then the code must explicitly execute that workflow and record those quantities at the required points.
- Do not access the network, browse the web, call external APIs or services, or otherwise rely on internet connectivity during the run.
- Do not use online resources, hidden baselines, copied reference solutions, or task-specific hints.
- Do not delegate any part of the task to sub-agents, helper agents, external agents, or parallel agent workers. Complete the work within the single provided agent session.
- The submitted `solution_N.py` should be concise. More than 200 non-empty, non-comment Python lines is considered a failed implementation strategy.
- The submitted solution must complete the evaluator's timed `run_solution(config)` call within 300 seconds. Runtime starts losing score after 180 seconds and receives zero runtime score at 300 seconds or above.
- Optimize for end-to-end evaluator runtime, not just correctness. Prefer the most direct and efficient {framework} representation, reduce avoidable Python overhead, reuse compiled or precomputed objects when appropriate, and iterate on the implementation if a simpler first attempt is too slow.
- When using a JAX-backed implementation, pay attention not only to execution speed but also to compilation latency. When appropriate, prefer structured scans such as `jax.scan` over large Python-unrolled update loops or deeply unrolled layer-by-layer circuit construction, since scanning repeated circuit blocks or optimizer steps can materially reduce compile overhead.
- When there are multiple plausible approaches, test or reason about which one is likely fastest within {framework}, then use that path. Avoid spending time polishing slow designs that are unlikely to pass the runtime budget.
- Before committing to a deep implementation path, verify that the installed {framework} and its dependencies actually expose the primitives, APIs, and numerical behavior the task needs.
- Stop early only if, after checking realistic framework-native alternatives, you conclude that solving the task with the required framework, allowed methods, and acceptable runtime is genuinely close to impossible.
- If your investigation shows that the task is very unlikely to fit within the allowed runtime using the required framework and allowed methods, stop early and report that efficiency obstruction instead of spending the remaining budget on low-probability tuning attempts.

Before implementing, inspect the installed {framework} package, examples, and source code available in the task environment. Search for APIs, tensor-network primitives, circuit/state representations, measurement utilities, optimizers, and interoperability hooks that make the task feasible and efficient in {framework}. Choose an implementation strategy based on what the framework can actually express well.

If you find a substantive obstruction that makes the task impractical or impossible to solve within the required framework and runtime budget, declare the task failed clearly in your final response and stop work promptly rather than spending the entire run on an unproductive workaround. Do not bypass the obstruction by switching frameworks, by importing or using another quantum framework, by writing a raw quantum simulator from scratch, by reverse-engineering the evaluator thresholds, or by manufacturing outputs that mimic a successful run without actually performing the required computation.
\end{lstlisting}

\section*{Supplementary Note 9: TC performance-checklist prompt}

The intervention added the TC-specific workflow block shown in Supplementary Listing~\ref{lst:tc-checklist-prompt} after the framework-constraint prompt in Supplementary Listing~\ref{lst:framework-prompt}.
It was a generic efficiency scaffold and did not contain task-specific answers.

\begin{lstlisting}[style=promptlisting,caption={TC-specific performance-checklist prompt block.},label={lst:tc-checklist-prompt}]
## Time Budget

You have only 30 minutes of total solve time in this agent run. Act with urgency. Use tools, code inspection, and tightly scoped test snippets quickly and intelligently. Prioritize early detection of viable TensorCircuit-native solution paths, fast elimination of bad designs, and short empirical checks over long speculative implementation detours.

Preserve a runnable answer as early as possible. Once you have a minimally working solution that plausibly satisfies the task contract, keep that version intact and iterate from there with small, testable improvements rather than risking the whole run on a late large rewrite.

## TensorCircuit Performance Checklist

Use the following checklist as a source of optimization hypotheses and implementation ideas. These are not automatic rules. You must judge them against the concrete task.

- Prefer the most native representation that matches the physics: use TensorCircuit-native operators and abstractions when they fit, including options such as `tc.quantum.PauliStringSum2MVP`, `tc.quantum.PauliStringSum2COO`, `tc.templates.measurements.mpo_expectation`, direct MPS/MPO/qop contraction, `Circuit.post_select`, `enable_lightcone=True`, `QuditCircuit`, and built-in multi-qubit gates such as `cmz` or `su4`.
- Vectorization: replace manual Python loops with `tc.backend.vmap` or `jax.vmap` when this reduces overhead.
- JIT compilation: wrap performance-critical tensor-in/tensor-out kernels with `tc.backend.jit` or `jax.jit`, preferably around the outer optimization step rather than tiny inner fragments. Remember that JIT can be a net loss if the function only runs once.
- JIT reuse discipline: keep input shapes, dtypes, Python container structure, and static arguments stable across calls. Put structural settings in static args when needed, and do not change dtype or shape mid-loop if you want compilation reuse.
- JIT boundary for gradients and optimizers: often prefer jitting the whole train step instead of returning raw gradients from a jitted `value_and_grad` and then updating in Python.
- Move static physics and fixed data out of the hot path: prebuild Hamiltonians, MPOs, matrix-free operators, target states, fixed gate tapes, bond lists, sampled Pauli structures, and reproducible shot-status tensors once and reuse them.
- Structured parameters over flat vectors: prefer PyTrees or structured parameter blocks over a single large flattened parameter vector when that avoids costly slice-based autodiff and gradient assembly.
- Use locality-aware evaluation early: if the task only needs local observables or sparse measurements, test whether light-cone reduction such as `enable_lightcone=True` cuts work materially.
- XLA fusion pathology check: if first-call compile time is unexpectedly extreme, test whether XLA fusion/codegen is the real bottleneck rather than TensorCircuit tracing.
- Staging awareness: single-qubit-heavy unrolling can have worse staging behavior than expected, so avoid gratuitous unrolling.
- Static setup vs numeric loop: move setup work out of the timed loop, but distinguish one-time setup savings from real steady-state speedups.
- Sampling path choice matters: for large-qubit or large-shot cases, benchmark direct sampling paths such as `Circuit.sample(batch=..., allow_state=False)` when the explicit final state is unnecessary.
- Host-device transfer hygiene: avoid `.item()`, Python `float(...)`, `tc.backend.numpy(...)`, printing tensors, or Python branching on tensors inside jitted or repeatedly executed kernels.
- Dtype policy: default to `complex64` and `float32` unless the task genuinely needs higher precision. Set dtype before building long-lived constants and before tracing.
- Avoid full state instantiation when unnecessary: for larger systems, test whether settings such as `reuse=False` or `allow_state=False` prevent wasteful dense-state materialization.
- Control flow, memory, and checkpointing: for repeated blocks, consider `jax.lax.scan` to reduce graph size, and use `jax.checkpoint` or `jax.remat` only when the memory-time trade-off is favorable in practice.
- Compile-time mode vs throughput mode: treat `scan`, partial unrolling, and full unrolling as different performance modes. `scan` may lower compile cost, while manual unrolling may still win on steady-state runtime.
- Contraction optimization: for large tensor-network workloads, test contractor choice and path-search budget. OMECO is a strong early candidate, and tuned cotengra may win when the path is heavily amortized.
- Contraction slicing or hybrid workflows: if exact contraction remains too expensive, consider an unsliced strong path first and then sliced or reconfigured execution when the code path actually supports it.
- Contraction path assumption: do not assume path search is the repeated bottleneck inside a jitted workload unless profiling shows retracing or changing circuit structure.
- Contraction tuning is not always the first lever: first test representation improvements such as light-cone reduction, sparse or MPO reformulation, direct overlap, or more native operator APIs before spending a lot of effort on path search.
- Sparse, MPO, and matrix-free observables or ODEs: for large Hamiltonians or Pauli sums, benchmark sparse or matrix-free routes such as `tc.templates.measurements.sparse_expectation`, `tc.templates.measurements.mpo_expectation`, `tc.quantum.PauliStringSum2MVP`, or `tc.quantum.PauliStringSum2COO` instead of dense assembly or manual `kron`.
- Backend selection and interfacing: if the task leaves room for backend choice within TensorCircuit, JAX is usually the best first option for speed, JIT, and vectorization. If a larger PyTorch stack is required, consider bridging via `tc.interfaces.torch_interface`.

These checklist items must be evaluated by real testing on the concrete workload. Do not assume any item is inherently positive. A change can improve runtime, worsen compile latency, increase memory, or do the opposite. Use short A/B checks whenever feasible, separate one-time setup or warmup cost from steady-state execution cost, and keep only the changes that empirically help this task.
\end{lstlisting}

The checklist preserved the same TC success count while reducing agent-side resource use; the artifact runtime was nearly unchanged.

\section*{Supplementary Note 10: Agent-level task map on TC}

\begin{figure}[htbp]
\centering
\includegraphics[width=\textwidth]{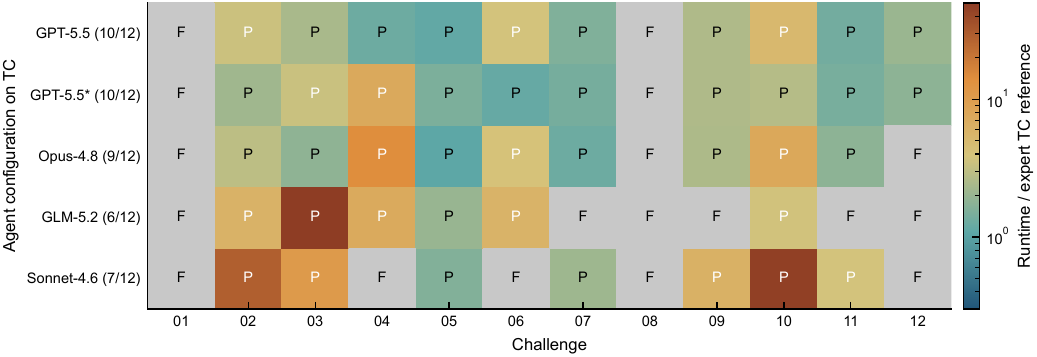}
\caption{\textbf{Task-level model comparison on TC.} Each cell corresponds to one challenge task under a fixed TC framework constraint. Passed tasks are marked P and colored by runtime relative to the expert TC reference. Failed tasks are marked F. The row labels report the final number of valid solutions for each model configuration. GPT-5.5 uses the Codex agent. Opus-4.8, GLM-5.2, and Sonnet-4.6 use the Claude Code agent. GPT-5.5* denotes GPT-5.5 with the TC-specific performance-checklist prompt.}
\label{fig:supp-tc-agent-task-map}
\end{figure}

Supplementary Fig.~\ref{fig:supp-tc-agent-task-map} shows the task-level validity and runtime pattern for the TC agent-axis comparison.
Each row is one agent configuration evaluated on TC.
The row label gives the final success count over 12 tasks.
For passed tasks, color encodes the runtime relative to the expert TC reference for the same task; failed tasks are marked F using a separate neutral color.

\section*{Supplementary Note 11: Complete benchmark results}

Supplementary Tables~\ref{tab:all-config-summary}--\ref{tab:details-mindquantum-codex} report the full experimental benchmark results used in the main text.
Agent-side costs exclude verifier-side semantic-audit usage.
Abbreviations used in these tables: VQE, variational quantum eigensolver; MPS, matrix-product state; DMRG, density-matrix renormalization group; RDM, reduced density matrix; TQ, TorchQuantum; TN, tensor network; CMZ, controlled multi-qubit Z rotation; MQ, MindQuantum.

\begin{table}[htbp]
\centering
\scriptsize
\caption{\textbf{Configuration-level benchmark summary.} Success counts, agent-side wall time, token use, and cost are aggregated over all 12 tasks in each configuration.}
\label{tab:all-config-summary}
\begin{tabular}{lrrrrrr}
\toprule
Configuration & Passes & Failures & Agent wall & Tokens & Cost & Cost/pass \\
\midrule
TC Codex (GPT-5.5) & 10 & 2 & 2h 5m 12.3s & 14,241,691 & \$17.99 & \$1.80 \\
TC Codex (GPT-5.5*) & 10 & 2 & 1h 29m 3.4s & 8,205,577 & \$13.29 & \$1.33 \\
TC Claude Code (Claude Opus-4.8) & 9 & 3 & 2h 53m 2.4s & 12,203,758 & \$17.52 & \$1.95 \\
TC Claude Code (GLM-5.2) & 6 & 6 & 3h 50m 11.1s & 21,075,177 & \$12.98 & \$2.16 \\
TC Claude Code (Claude Sonnet-4.6) & 7 & 5 & 3h 47m 24.2s & 21,268,795 & \$14.55 & \$2.08 \\
PennyLane Codex (GPT-5.5) & 8 & 4 & 2h 39m 35.7s & 16,579,728 & \$23.04 & \$2.88 \\
TorchQuantum Codex (GPT-5.5) & 4 & 8 & 2h 18m 30.5s & 20,298,642 & \$24.91 & \$6.23 \\
MindQuantum Codex (GPT-5.5) & 4 & 8 & 2h 42m 42.4s & 22,928,678 & \$29.03 & \$7.26 \\
\bottomrule
\end{tabular}
\end{table}

\clearpage
\scriptsize
\setlength{\tabcolsep}{2pt}
\begin{longtable}{@{}L{0.040\linewidth}L{0.040\linewidth}L{0.080\linewidth}L{0.080\linewidth}L{0.085\linewidth}L{0.085\linewidth}L{0.060\linewidth}L{0.415\linewidth}@{}}
\caption{\textbf{Task-level benchmark details for TC Codex (GPT-5.5).} Each row corresponds to one challenge task.}\label{tab:details-tc-codex}\\
\toprule
Task & Pass & Solution runtime & Reference runtime & Agent solve time & Agent tokens & Agent cost & Notes \\
\midrule
\endfirsthead
\caption[]{\textbf{Task-level benchmark details for TC Codex (GPT-5.5) continued.}}\\
\toprule
Task & Pass & Solution runtime & Reference runtime & Agent solve time & Agent tokens & Agent cost & Notes \\
\midrule
\endhead
01 & No & n/a & 27.22s & 28m 35.5s & 4,196,140 & \$3.59 & Agent timed out before producing a scorable artifact \\
02 & Yes & 9.90s & 2.87s & 5m 53.3s & 1,031,334 & \$1.39 & \\
03 & Yes & 5.99s & 2.46s & 3m 7.1s & 494,792 & \$1.02 & \\
04 & Yes & 15.19s & 11.83s & 15m 29.1s & 1,390,826 & \$1.63 & \\
05 & Yes & 51.56s & 45.50s & 15m 46.5s & 1,200,716 & \$1.69 & \\
06 & Yes & 1m 40.3s & 26.83s & 9m 44.9s & 1,271,536 & \$1.57 & \\
07 & Yes & 1m 38.9s & 1m 3.8s & 7m 38.6s & 502,176 & \$1.05 & \\
08 & No & 5.43s & 25.05s & 11m 20.4s & 1,422,944 & \$1.92 & Functional pass, semantic-audit failure: custom contraction/sampling bypassed TC \\
09 & Yes & 35.61s & 13.74s & 4m 37.4s & 277,267 & \$0.71 & \\
10 & Yes & 1m 2.8s & 12.44s & 6m 50.6s & 460,817 & \$0.75 & \\
11 & Yes & 1m 32.7s & 1m 8.1s & 9m 17.0s & 1,119,151 & \$1.49 & \\
12 & Yes & 12.74s & 6.12s & 6m 51.9s & 873,992 & \$1.18 & \\
\bottomrule
\end{longtable}
\normalsize

\scriptsize
\setlength{\tabcolsep}{2pt}
\begin{longtable}{@{}L{0.040\linewidth}L{0.040\linewidth}L{0.080\linewidth}L{0.080\linewidth}L{0.085\linewidth}L{0.085\linewidth}L{0.060\linewidth}L{0.415\linewidth}@{}}
\caption{\textbf{Task-level benchmark details for TC Codex (GPT-5.5*).} Each row corresponds to one challenge task.}\label{tab:details-tc-codex-plus}\\
\toprule
Task & Pass & Solution runtime & Reference runtime & Agent solve time & Agent tokens & Agent cost & Notes \\
\midrule
\endfirsthead
\caption[]{\textbf{Task-level benchmark details for TC Codex (GPT-5.5*) continued.}}\\
\toprule
Task & Pass & Solution runtime & Reference runtime & Agent solve time & Agent tokens & Agent cost & Notes \\
\midrule
\endhead
01 & No & 0.36s & 27.22s & 12m 58.3s & 1,731,676 & \$2.32 & Functional pass, semantic-audit failure: simplified local-rotation model replaces brickwork VQE \\
02 & Yes & 6.35s & 2.87s & 4m 44.2s & 481,893 & \$0.85 & \\
03 & Yes & 8.38s & 2.46s & 4m 27.7s & 449,821 & \$0.86 & \\
04 & Yes & 1m 26.6s & 11.83s & 9m 53.9s & 596,821 & \$0.98 & \\
05 & Yes & 1m 8.5s & 45.50s & 10m 25.3s & 960,110 & \$1.42 & \\
06 & Yes & 31.38s & 26.83s & 3m 38.4s & 330,129 & \$0.83 & \\
07 & Yes & 1m 30.7s & 1m 3.8s & 4m 21.7s & 339,750 & \$0.62 & \\
08 & No & 8.86s & 25.05s & 10m 47.3s & 928,227 & \$1.75 & Human-review failure: custom NumPy column-transfer contraction/sampling bypassed TC \\
09 & Yes & 35.38s & 13.74s & 3m 40.8s & 224,685 & \$0.50 & \\
10 & Yes & 34.18s & 12.44s & 3m 33.6s & 428,818 & \$0.68 & \\
11 & Yes & 1m 38.7s & 1m 8.1s & 14m 10.2s & 791,903 & \$1.27 & \\
12 & Yes & 11.00s & 6.12s & 6m 22.0s & 941,744 & \$1.21 & \\
\bottomrule
\end{longtable}
\normalsize

\clearpage
\scriptsize
\setlength{\tabcolsep}{2pt}
\begin{longtable}{@{}L{0.040\linewidth}L{0.040\linewidth}L{0.080\linewidth}L{0.080\linewidth}L{0.085\linewidth}L{0.085\linewidth}L{0.060\linewidth}L{0.415\linewidth}@{}}
\caption{\textbf{Task-level benchmark details for TC Claude Code (Claude Opus-4.8).} Each row corresponds to one challenge task.}\label{tab:details-tc-opus}\\
\toprule
Task & Pass & Solution runtime & Reference runtime & Agent solve time & Agent tokens & Agent cost & Notes \\
\midrule
\endfirsthead
\caption[]{\textbf{Task-level benchmark details for TC Claude Code (Claude Opus-4.8) continued.}}\\
\toprule
Task & Pass & Solution runtime & Reference runtime & Agent solve time & Agent tokens & Agent cost & Notes \\
\midrule
\endhead
01 & No & n/a & 27.22s & 21m 45.5s & 2,784,543 & \$2.28 & Submitted artifact timed out during  evaluation \\
02 & Yes & 8.50s & 2.87s & 6m 24.2s & 810,473 & \$1.29 & \\
03 & Yes & 4.54s & 2.46s & 23m 35.1s & 1,079,876 & \$1.51 & \\
04 & Yes & 2m 47.0s & 11.83s & 30m 0.0s & 1,524,732 & \$2.62 & \\
05 & Yes & 48.36s & 45.50s & 9m 18.2s & 779,913 & \$1.34 & \\
06 & Yes & 1m 47.1s & 26.83s & 9m 13.1s & 743,764 & \$1.37 & \\
07 & Yes & 1m 21.4s & 1m 3.8s & 30m 0.0s & 1,156,796 & \$2.08 & \\
08 & No & n/a & 25.05s & 2m 31.8s & 194,149 & \$0.51 & Cyber-safeguard refusal before solution was produced \\
09 & Yes & 34.41s & 13.74s & 8m 50.3s & 551,429 & \$0.97 & \\
10 & Yes & 1m 33.3s & 12.44s & 9m 21.0s & 741,753 & \$0.78 & \\
11 & Yes & 2m 2.0s & 1m 8.1s & 19m 10.2s & 1,523,481 & \$1.96 & \\
12 & No & n/a & 6.12s & 2m 53.0s & 312,849 & \$0.81 & Cyber-safeguard refusal before solution was produced \\
\bottomrule
\end{longtable}
\normalsize

\scriptsize
\setlength{\tabcolsep}{2pt}
\begin{longtable}{@{}L{0.040\linewidth}L{0.040\linewidth}L{0.080\linewidth}L{0.080\linewidth}L{0.085\linewidth}L{0.085\linewidth}L{0.060\linewidth}L{0.415\linewidth}@{}}
\caption{\textbf{Task-level benchmark details for TC Claude Code (GLM-5.2).} Each row corresponds to one challenge task.}\label{tab:details-tc-glm}\\
\toprule
Task & Pass & Solution runtime & Reference runtime & Agent solve time & Agent tokens & Agent cost & Notes \\
\midrule
\endfirsthead
\caption[]{\textbf{Task-level benchmark details for TC Claude Code (GLM-5.2) continued.}}\\
\toprule
Task & Pass & Solution runtime & Reference runtime & Agent solve time & Agent tokens & Agent cost & Notes \\
\midrule
\endhead
01 & No & n/a & 27.22s & 30m 0.1s & 3,680,174 & \$3.19 & Agent timed out before producing a solution artifact \\
02 & Yes & 16.89s & 2.87s & 5m 2.0s & 570,510 & \$0.38 & \\
03 & Yes & 1m 56.7s & 2.46s & 30m 0.1s & 3,518,912 & \$2.48 & \\
04 & Yes & 1m 23.8s & 11.83s & 24m 9.8s & 2,540,761 & \$1.07 & \\
05 & Yes & 1m 32.3s & 45.50s & 11m 59.3s & 592,114 & \$0.37 & \\
06 & Yes & 2m 32.8s & 26.83s & 25m 24.0s & 2,204,624 & \$1.42 & \\
07 & No & n/a & 1m 3.8s & 24m 34.1s & 1,225,285 & \$0.73 & Agent exited 137 after heavy TC DMCircuit running; no scorable artifact \\
08 & No & n/a & 25.05s & 12m 0.4s & 555,137 & \$0.35 & Agent exited 137 after heavy TC contraction/sampling running; no scorable artifact \\
09 & No & 8.47s & 13.74s & 10m 48.1s & 600,180 & \$0.34 & Functional pass, semantic-audit failure: omitted required plus-state initialization \\
10 & Yes & 45.67s & 12.44s & 17m 19.0s & 1,738,141 & \$0.79 & \\
11 & No & n/a & 1m 8.1s & 21m 33.5s & 1,530,951 & \$0.79 & Functional evaluation stage killed after resource-heavy qutrit execution \\
12 & No & 2m 29.7s & 6.12s & 17m 20.7s & 2,318,388 & \$1.07 & Functional pass, semantic-audit failure: mismatched implementation from the problem \\
\bottomrule
\end{longtable}
\normalsize

\clearpage
\scriptsize
\setlength{\tabcolsep}{2pt}
\begin{longtable}{@{}L{0.040\linewidth}L{0.040\linewidth}L{0.080\linewidth}L{0.080\linewidth}L{0.085\linewidth}L{0.085\linewidth}L{0.060\linewidth}L{0.415\linewidth}@{}}
\caption{\textbf{Task-level benchmark details for TC Claude Code (Claude Sonnet-4.6).} Each row corresponds to one challenge task.}\label{tab:details-tc-sonnet}\\
\toprule
Task & Pass & Solution runtime & Reference runtime & Agent solve time & Agent tokens & Agent cost & Notes \\
\midrule
\endfirsthead
\caption[]{\textbf{Task-level benchmark details for TC Claude Code (Claude Sonnet-4.6) continued.}}\\
\toprule
Task & Pass & Solution runtime & Reference runtime & Agent solve time & Agent tokens & Agent cost & Notes \\
\midrule
\endhead
01 & No & n/a & 27.22s & 30m 0.0s & 4,351,656 & \$2.01 & Submitted artifact timed out during functional evaluation \\
02 & Yes & 1m 21.6s & 2.87s & 30m 0.0s & 2,177,375 & \$2.05 & \\
03 & Yes & 25.67s & 2.46s & 24m 56.9s & 2,346,526 & \$1.59 & \\
04 & No & n/a & 11.83s & 30m 0.1s & 2,772,013 & \$2.31 & Agent timed out before producing a solution artifact \\
05 & Yes & 1m 12.0s & 45.50s & 11m 43.6s & 1,369,687 & \$1.15 & \\
06 & No & n/a & 26.83s & 5m 37.5s & 610,159 & \$0.44 & Runtime API exception: PauliStringSum2COO received string Pauli labels under JAX \\
07 & Yes & 2m 17.2s & 1m 3.8s & 18m 53.8s & 1,591,490 & \$0.79 & \\
08 & No & 2.87s & 25.05s & 19m 9.5s & 1,311,477 & \$1.13 & Functional pass, semantic-audit failure: handwritten 2D tensor-network sampling bypassing TC \\
09 & Yes & 1m 22.9s & 13.74s & 12m 56.3s & 1,191,784 & \$0.78 & \\
10 & Yes & 9m 27.7s & 12.44s & 7m 40.0s & 959,901 & \$0.53 & \\
11 & Yes & 4m 16.9s & 1m 8.1s & 22m 24.4s & 1,914,092 & \$0.88 & \\
12 & No & n/a & 6.12s & 14m 2.1s & 672,635 & \$0.89 & No solution artifact was produced \\
\bottomrule
\end{longtable}
\normalsize

\scriptsize
\setlength{\tabcolsep}{2pt}
\begin{longtable}{@{}L{0.040\linewidth}L{0.040\linewidth}L{0.080\linewidth}L{0.080\linewidth}L{0.085\linewidth}L{0.085\linewidth}L{0.060\linewidth}L{0.415\linewidth}@{}}
\caption{\textbf{Task-level benchmark details for PennyLane Codex (GPT-5.5).} Each row corresponds to one challenge task.}\label{tab:details-pennylane-codex}\\
\toprule
Task & Pass & Solution runtime & Reference runtime & Agent solve time & Agent tokens & Agent cost & Notes \\
\midrule
\endfirsthead
\caption[]{\textbf{Task-level benchmark details for PennyLane Codex (GPT-5.5) continued.}}\\
\toprule
Task & Pass & Solution runtime & Reference runtime & Agent solve time & Agent tokens & Agent cost & Notes \\
\midrule
\endhead
01 & No & 3m 3.9s & 27.22s & 19m 44.2s & 2,361,150 & \$3.02 & Semantic-audit failure: sparse/partial optimization of declared ansatz \\
02 & Yes & 9.95s & 2.87s & 3m 13.0s & 207,104 & \$0.67 & \\
03 & Yes & 38.95s & 2.46s & 4m 11.6s & 608,661 & \$0.95 & \\
04 & No & 8.12s & 11.83s & 7m 49.3s & 579,648 & \$1.30 & Semantic-audit failure: hand-written analytic simulator replaces PennyLane noisy circuit \\
05 & Yes & 2m 11.8s & 45.50s & 10m 55.1s & 895,907 & \$1.22 & \\
06 & Yes & 1m 39.7s & 26.83s & 8m 25.0s & 1,235,168 & \$1.50 & \\
07 & No & 5.54s & 1m 3.8s & 10m 14.4s & 1,281,060 & \$1.83 & Semantic-audit failure: trains a surrogate objective \\
08 & Yes & 1m 0.0s & 25.05s & 17m 20.5s & 1,805,137 & \$2.52 & \\
09 & Yes & 2m 41.0s & 13.74s & 29m 26.7s & 2,412,391 & \$3.61 & \\
10 & Yes & 3m 52.9s & 12.44s & 18m 30.7s & 1,716,467 & \$2.13 & \\
11 & Yes & 2m 49.1s & 1m 8.1s & 15m 59.9s & 2,165,471 & \$2.47 & \\
12 & No & 3.25s & 6.12s & 13m 45.3s & 1,311,564 & \$1.82 & Semantic-audit failure: hand-written JAX contraction bypassing PennyLane \\
\bottomrule
\end{longtable}
\normalsize

\clearpage
\scriptsize
\setlength{\tabcolsep}{2pt}
\begin{longtable}{@{}L{0.040\linewidth}L{0.040\linewidth}L{0.080\linewidth}L{0.080\linewidth}L{0.085\linewidth}L{0.085\linewidth}L{0.060\linewidth}L{0.415\linewidth}@{}}
\caption{\textbf{Task-level benchmark details for TorchQuantum Codex (GPT-5.5).} Each row corresponds to one challenge task.}\label{tab:details-torchquantum-codex}\\
\toprule
Task & Pass & Solution runtime & Reference runtime & Agent solve time & Agent tokens & Agent cost & Notes \\
\midrule
\endfirsthead
\caption[]{\textbf{Task-level benchmark details for TorchQuantum Codex (GPT-5.5) continued.}}\\
\toprule
Task & Pass & Solution runtime & Reference runtime & Agent solve time & Agent tokens & Agent cost & Notes \\
\midrule
\endhead
01 & No & 3.90s & 27.22s & 12m 7.3s & 2,643,499 & \$3.10 & Functional pass, semantic-audit failure: local-RDM surrogate replaces full refinement circuit \\
02 & Yes & 27.04s & 2.87s & 5m 50.0s & 859,854 & \$1.45 & \\
03 & Yes & 21.07s & 2.46s & 5m 7.7s & 556,044 & \$0.98 & \\
04 & No & 5.38s & 11.83s & 11m 0.0s & 2,183,300 & \$2.62 & Functional pass, semantic-audit failure: hand-written Kraus bypassing TQ \\
05 & Yes & 1m 44.9s & 45.50s & 9m 46.6s & 1,773,601 & \$2.10 & \\
06 & Yes & 1m 43.6s & 26.83s & 11m 24.7s & 1,748,405 & \$2.14 & \\
07 & No & 3.37s & 1m 3.8s & 12m 31.5s & 1,670,437 & \$1.95 & Functional pass, semantic-audit failure: 8-qubit surrogate omits ancilla and feedback \\
08 & No & 8.38s & 25.05s & 13m 0.2s & 1,283,844 & \$1.86 & Human-review failure: custom TN/MPS sampler bypassing TQ \\
09 & No & 7.85s & 13.74s & 14m 19.2s & 2,343,141 & \$2.55 & Human-review failure: local-cone contraction bypassing TQ \\
10 & No & 3m 56.2s & 12.44s & 17m 55.7s & 1,768,361 & \$2.48 & Functional pass, semantic-audit failure: CMZ applied by raw state-tensor phases \\
11 & No & 2m 43.4s & 1m 8.1s & 14m 48.3s & 1,407,749 & \$1.54 & Functional pass, semantic-audit failure: qutrit rotation order mismatch \\
12 & No & 40.87s & 6.12s & 10m 39.3s & 2,060,407 & \$2.14 & Human-review failure: custom MPS contraction bypassing TQ \\
\bottomrule
\end{longtable}
\normalsize

\scriptsize
\setlength{\tabcolsep}{2pt}
\begin{longtable}{@{}L{0.040\linewidth}L{0.040\linewidth}L{0.080\linewidth}L{0.080\linewidth}L{0.085\linewidth}L{0.085\linewidth}L{0.060\linewidth}L{0.415\linewidth}@{}}
\caption{\textbf{Task-level benchmark details for MindQuantum Codex (GPT-5.5).} Each row corresponds to one challenge task.}\label{tab:details-mindquantum-codex}\\
\toprule
Task & Pass & Solution runtime & Reference runtime & Agent solve time & Agent tokens & Agent cost & Notes \\
\midrule
\endfirsthead
\caption[]{\textbf{Task-level benchmark details for MindQuantum Codex (GPT-5.5) continued.}}\\
\toprule
Task & Pass & Solution runtime & Reference runtime & Agent solve time & Agent tokens & Agent cost & Notes \\
\midrule
\endhead
01 & No & n/a & 27.22s & 4m 16.0s & 795,609 & \$1.54 & Declared obstruction: mqmps cannot support the required external-MPS workflow \\
02 & Yes & 9.94s & 2.87s & 4m 48.7s & 551,949 & \$1.17 & \\
03 & Yes & 53.66s & 2.46s & 9m 55.1s & 1,451,151 & \$2.14 & \\
04 & No & 7.67s & 11.83s & 6m 46.3s & 656,613 & \$1.53 & Functional pass, semantic-audit failure: hand-written NumPy/Kraus transfer simulation \\
05 & No & 1m 28.3s & 45.50s & 25m 43.1s & 5,073,450 & \$5.54 & Static/semantic failure: full NumPy statevector simulator, no MindQuantum import \\
06 & No & 2m 24.7s & 26.83s & 26m 13.6s & 2,091,062 & \$2.58 & Functional pass, semantic-audit failure: clipped parameters and finite-difference gradients \\
07 & No & 1m 18.6s & 1m 3.8s & 15m 25.1s & 1,683,461 & \$1.93 & Functional pass, semantic-audit failure: surrogate gradient replaces measurement-feedback objective \\
08 & No & 1.19s & 25.05s & 5m 38.6s & 1,158,466 & \$1.84 & Functional pass, semantic-audit failure: mqmps MPS route replaces 2D TN contraction \\
09 & Yes & 4m 49.4s & 13.74s & 28m 45.1s & 2,566,917 & \$2.99 & \\
10 & Yes & 1m 21.3s & 12.44s & 6m 59.3s & 446,389 & \$0.68 & \\
11 & No & n/a & 1m 8.1s & 19m 35.1s & 4,734,997 & \$4.69 & Missing solution artifact \\
12 & No & 1m 8.3s & 6.12s & 8m 36.4s & 1,718,614 & \$2.40 & Human-review failure: custom NumPy TN overlap/gradient bypassing MQ \\
\bottomrule
\end{longtable}
\normalsize